\documentclass[%
 reprint,
nofootinbib,
 amsmath,amssymb,
 aps,
]{revtex4-2}

\usepackage{graphicx}
\usepackage{dcolumn}
\usepackage{bm}
\usepackage[hypertexnames=false]{hyperref}
\usepackage[all]{hypcap}    
\usepackage{siunitx}
\usepackage{braket}
\usepackage{rotating}
\usepackage{xfrac}

\begin{document}

\title{Realization of a multi-node quantum network of remote solid-state qubits}

\author{M. Pompili}\altaffiliation{These authors contributed equally to this work}
\author{S. L. N. Hermans}\altaffiliation{These authors contributed equally to this work}
\author{S. Baier}\altaffiliation{These authors contributed equally to this work}
\affiliation{%
\vspace{0.5em}QuTech \& Kavli Institute of Nanoscience, Delft University of Technology, 2628 CJ Delft, The Netherlands
}
\author{H. K. C. Beukers}
\author{P. C. Humphreys}
\author{R. N. Schouten}
\author{R. F. L. Vermeulen}
\author{M. J. Tiggelman}
\author{L. dos Santos Martins}
\author{B. Dirkse}
\author{S. Wehner}
\author{R. Hanson}
\email{Correspondence to: R.Hanson@tudelft.nl}

\affiliation{%
\vspace{0.5em}QuTech \& Kavli Institute of Nanoscience, Delft University of Technology, 2628 CJ Delft, The Netherlands
}


\begin{abstract}
The distribution of entangled states across the nodes of a future quantum internet will unlock fundamentally new technologies. Here we report on the experimental realization of a three-node entanglement-based quantum network. We combine remote quantum nodes based on diamond communication qubits into a scalable phase-stabilized architecture, supplemented with a robust memory qubit and local quantum logic. In addition, we achieve real-time communication and feed-forward gate operations across the network. We capitalize on the novel capabilities of this network to realize two canonical protocols without post-selection: the distribution of genuine multipartite entangled states across the three nodes and entanglement swapping through an intermediary node. Our work establishes a key platform for exploring, testing and developing multi-node quantum network protocols and a quantum network control stack.
\end{abstract}

\footnotetext{S. B. is currently at: Institut für Experimentalphysik, Universität Innsbruck, Technikerstraße 25, 6020 Innsbruck, Austria}
\footnotetext{P. C. H. is currently at: DeepMind, London, UK}
\footnotetext{M. J. T. is currently at: QBlox, 2628 CJ Delft, The Netherlands}
\maketitle

Future quantum networks sharing entanglement across multiple nodes~\cite{kimble_quantum_2008, wehner_quantum_2018} promise to enable a range of novel applications such as secure communication, distributed quantum computing, enhanced sensing and fundamental tests of quantum mechanics~\cite{jiang_distributed_2007, broadbent_universal_2009, gottesman_longer-baseline_2012, ekert_ultimate_2014, nickerson_freely_2014, komar_quantum_2014}. Intensive research efforts in the past decade have focused on realizing the building blocks of such a network: quantum nodes capable of establishing remote entangled links as well as locally storing, processing and reading out quantum information.

Entanglement generation via optical channels between a pair of individually controlled qubits has been demonstrated with trapped ions and atoms~\cite{moehring_entanglement_2007, ritter_elementary_2012, hofmann_heralded_2012, stephenson_high-rate_2020}, diamond Nitrogen-Vacancy (NV) centers~\cite{bernien_heralded_2013, humphreys_deterministic_2018} and quantum dots~\cite{delteil_generation_2016, stockill_phase-tuned_2017}. In addition, a number of quantum network primitives have been explored on these elementary two-node links, including non-local quantum gates~\cite{maunz_heralded_2009} and entanglement distillation~\cite{kalb_entanglement_2017}. Moving these qubit platforms beyond two-node experiments has so far remained an outstanding challenge due to the combination of several demanding requirements. Multiple high-performance quantum nodes are needed that include a communication qubit with an optical interface as well as an efficient memory qubit for storage and processing. Additionally, the individual entanglement links need to be embedded into a multi-node quantum network, requiring a scalable architecture and multi-node control protocols.

Here we report on the realization and integration of all elements of a multi-node quantum network: optically-mediated entanglement links connected through an extensible architecture, local memory qubit and quantum logic, real-time heralding and feed-forward operations. We demonstrate the full operation of the multi-node network by running two key quantum network protocols. First, we establish Greenberger-Horne-Zeilinger (GHZ) entangled states across the three nodes. Such distributed genuine multipartite entangled states are a key ingredient for many network applications~\cite{wehner_quantum_2018} such as anonymous transmission~\cite{christandl_quantum_2005}, secret sharing~\cite{hillery_quantum_1999}, leader election~\cite{ambainis_multiparty_2004} and clock stabilization~\cite{komar_quantum_2014}. Second, we perform entanglement swapping through an intermediary node, which is the central protocol for entanglement routing on a quantum network enabling any-to-any connectivity~\cite{briegel_quantum_1998, pant_routing_2019}. Thanks to efficient coherence protection on all qubits, combined with real-time feed-forward operations, we are able to realize these protocols in a heralded fashion, delivering the final states ready for further use. 

\begin{figure*}
\includegraphics{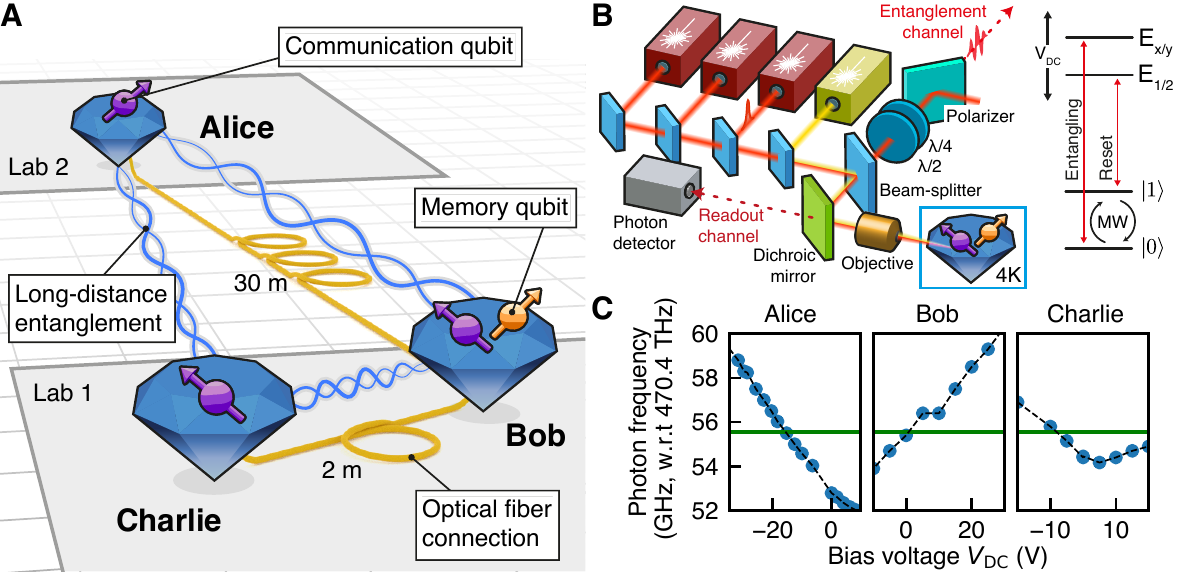}
\caption{\label{fig:network}The three-node quantum network. (A) Layout of the network. Three nodes, labelled Alice, Bob and Charlie, are located in two separate labs. Each node contains an NV center communication qubit (purple). At Bob, an additional nuclear spin qubit (orange) is used in the presented experiments. Fiber connections between the nodes (lengths indicated) enable remote entanglement generation on the links Alice-Bob and Bob-Charlie, which, combined with local quantum logic, allow for entanglement to be shared between all nodes (wiggly lines). (B) Left: simplified schematic of the optical setup at each node (see Fig.~\ref{fig:supp_optics}, Table~\ref{tab:supp_equipment} and Supplementary Material for additional details). Right: Diagram of the relevant levels of the electronic spin qubit, showing optical transitions for remote entanglement generation and readout (“entangling”), qubit reset (“reset”), and resonant microwaves (“MW”) for qubit control (see Figs.~\ref{fig:supp_levels},~\ref{fig:supp_skewness} for additional details). The memory qubit at Bob is initialized, controlled and read out via the electronic qubit (Fig.~\ref{fig:supp_nuclear-spin-control}). Optical transition frequencies are tuned via the DC bias voltages ($\mathrm{V}_\mathrm{DC}$). (C) Tuning of the optical “entangling” transition at each of the three nodes. Solid line is the working point, \SI{470.45555}{\tera\hertz}, dashed line is a guide to the eye.}
\end{figure*}

Our network is composed of three spatially separated quantum nodes (Fig.~\ref{fig:network}A-B), labelled Alice, Bob and Charlie. Each node consists of an NV center electronic spin as communication qubit. In addition, the middle node Bob employs a Carbon-13 nuclear spin as a memory qubit. Initialization and single-shot readout of the communication qubits are performed through resonant optical excitation and measurement of state-dependent fluorescence~\cite{humphreys_deterministic_2018}. Universal quantum logic on the electronic-nuclear register is achieved through tailored microwave pulses delivered on chip (see Supplementary Material). The nodes are connected through an optical fiber network for the quantum signals, as well as classical communication channels for synchronizing the control operations and relaying heralding signals (see below).

Remote entanglement generation hinges on indistinguishability between emitted photons. For NV centers in high-purity low-strain diamond devices, the optical transition frequencies show relatively minor variations (few GHz). We remove the remaining offsets by employing DC Stark tuning at each node via bias fields generated on chip (Fig.~\ref{fig:network}C). We are thus able to bring the relevant optical transitions of all three nodes to the same frequency, which we choose to be the zero-bias frequency of Bob.

\section{Establishing remote entanglement in a network architecture}
\begin{figure*}
\includegraphics{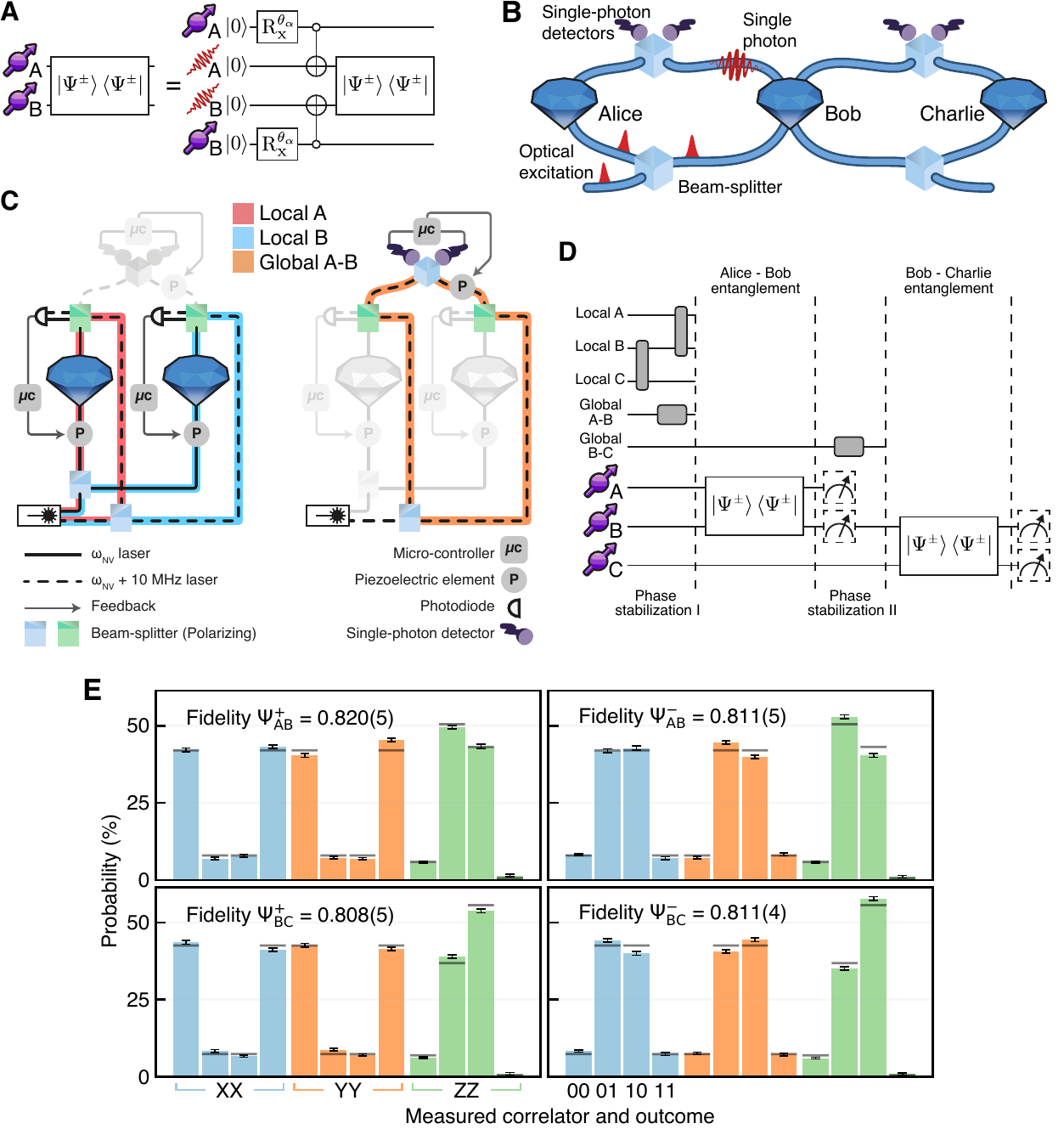}
\caption{\label{fig:phase}Establishing remote entanglement in a network architecture. (A) Circuit diagram of the single-photon entanglement protocol, where $\theta_\alpha = 2 \cos^{-1}\left(\sqrt{\alpha}\right)$ . (B) Sketch of three quantum network nodes in line configuration, showing the two effective interferometers. (C) Phase stabilization diagram of the Alice – Bob link, highlighting the local interferometers (left) and the global interferometer (right). See Supplementary Material for further details. (D) Experimental sequence to generate Bell pairs on both Alice-Bob (A-B) and Bob-Charlie (B-C) links. Dashed boxes display measurements used in (E). (E) Correlation measurements on entangled states on A-B (top) and B-C (bottom) links. Left (right) corresponds to $\ket{\Psi^+} \left( \ket{\Psi^-}\right)$  states. Shown are observed probabilities for outcomes (from left to right) 00, 01, 10 and 11 for correlation measurements in the bases XX (blue), YY (orange) and ZZ (green). Grey bars depict values from the theoretical model. Error bars indicate one standard deviation.}
\end{figure*}

To generate remote entanglement between a pair of nodes (i.e.~one elementary link) we employ a single-photon protocol~\cite{cabrillo_creation_1999, bose_proposal_1999}, of which the corresponding circuit diagram is depicted in Figure~\ref{fig:phase}A. The communication qubits of the nodes are each prepared in a superposition state $\ket\alpha = \sqrt\alpha \ket 0 + \sqrt{1 - \alpha} \ket 1$ . At each node, pulsed optical excitation, that is resonant only for the $\ket 0$ state, and subsequent photon emission deterministically create an entangled state between the communication qubit and the presence/absence of a photon (the flying qubit). The photonic modes from the two nodes are then interfered on a beam-splitter, removing the which-path information. The beam-splitter closes an effective interferometer formed by the optical excitation and collection paths. Detection of a single photon after the beam-splitter heralds the state $\ket{\psi^\pm}\approx \left( \ket{01} \pm e^{i\Delta\theta}\ket{10}\right)/\sqrt{2}$ between the two communication qubits, where the $\pm$ sign depends on which of the two detectors clicked and $\Delta \theta$ is the optical phase difference between the two arms of the effective interferometer (see Supplementary Material). Experimentally, this phase difference is set to a known value by stabilizing the full optical path using a feedback loop, as demonstrated in Refs.~\cite{stockill_phase-tuned_2017, humphreys_deterministic_2018}. This scheme yields states at maximum fidelity $1 - \alpha$ at a rate $\approx 2~\alpha~p_\mathrm{det}$, with $p_\mathrm{det}$ the probability that an emitted photon is detected.

Scaling this entangling scheme to multiple nodes requires each elementary link to be phase stabilized independently (Fig.~\ref{fig:phase}B), posing a number of new challenges. The different links, and even different segments of the same link, will generally be subject to diverse noise levels and spectra. Additionally, the optical power levels used are vastly different, from microwatts for the excitation path to attowatts for the single-photon heralding station, requiring different detector technologies for optimal signal detection. We solve these challenges with a hybrid phase stabilization scheme that is scalable to an arbitrary number of nodes. We decompose the effective interferometer for each link into three independently addressable interferometers and stabilize each separately (see Fig.~\ref{fig:phase}C for the Alice-Bob link; the link Bob-Charlie is phase-stabilized in an analogous and symmetric way, see Figs.~\ref{fig:supp_phase_diagram_total} to~\ref{fig:supp_phase_timings}). 

First, each node has its own local stabilization that uses unbalanced heterodyne phase detection (Fig.~\ref{fig:phase}C, left). In comparison to the previous homodyne stabilization method~\cite{humphreys_deterministic_2018}, this enables us to obtain a higher bandwidth phase signal from the small part of the excitation light that is reflected from the diamond surface ($\approx\SI{1}{\percent}$) by boosting it with a strong reference-light beam at a known frequency offset. Moreover, this scheme allows for optimal rejection of the reflected excitation light by polarization selection, thus preventing excitation light from entering the single-photon path towards the heralding detectors and creating false entanglement heralding events. The measured phase signals are fed back on piezoelectric-mounted mirrors to stabilize the local interferometers.

Second, the global part of the effective interferometer (Fig.~\ref{fig:phase}C, right) is stabilized by single-photon-level homodyne phase detection with feedback on a fiber stretcher: a small fraction of the strong reference-light beam is directed into the single-photon path and the interference is measured using the same detectors used for entanglement generation. 

This architecture provides scalability in the number of nodes, and a higher feedback bandwidth compared to our previous implementation on a single link (Fig.~\ref{fig:supp_phase_results}, see Supplementary Material for details). In our current implementation, the central node – Bob – has combining optics to merge the signals coming from Alice and Charlie, so that the single-photon detectors can be shared by the two links. 

Crucially, this architecture enables the successive generation of entanglement on the two elementary links as required for network protocols exploiting multi-node entanglement. We benchmark its performance by running entanglement generation on both elementary links within a single experimental sequence (Fig.~\ref{fig:phase}D). We achieve fidelities of the entangled Bell states exceeding \num{0.8} for both links (Fig.~\ref{fig:phase}E), on par with the highest fidelity reported for a single link in Ref.~\cite{humphreys_deterministic_2018}. Thanks to the new architecture, for the same fidelity the entangling rates are slightly higher than in Ref.~\cite{humphreys_deterministic_2018} (\SI{9}{\hertz} and \SI{7}{\hertz} for links Alice-Bob and Bob-Charlie, respectively), despite the additional channel loss from connecting the two links. The main sources of infidelity are the probability $\alpha$ that both nodes emit a photon, remaining optical phase uncertainty and double excitation during the optical pulse (see Table~\ref{tab:supp_budget_epr} and Supplementary Material). A detailed physical model including known error sources is used here and below for comparison to the experimental data (see Supplementary Material); predictions by the model are indicated by the grey bars in the correlation and fidelity plots.

\section{Memory qubit performance and real-time feed-forward operations}
In order to distribute entangled states across multiple nodes, generated entangled states must be stored in additional qubits while new entanglement links are created. Carbon-13 nuclear spins are excellent candidates for such memory qubits thanks to their long coherence times, controllability and isolation from the control drives on the electronic qubit~\cite{bradley_ten-qubit_2019}. Recent work~\cite{kalb_dephasing_2018} indicated that their storage fidelity under network activity is mainly limited by dephasing errors resulting from the coupling to the electronic spin that is randomized on failed entanglement generation. It was suggested that the memory robustness to such errors may be further improved by operating under an increased applied magnetic field. Here we use a magnetic field of \SI{189}{\milli\tesla} for our central node, as opposed to $\approx \SI{40}{\milli\tesla}$ used in past experiments~\cite{kalb_entanglement_2017, kalb_dephasing_2018}. 

\begin{figure}
\includegraphics{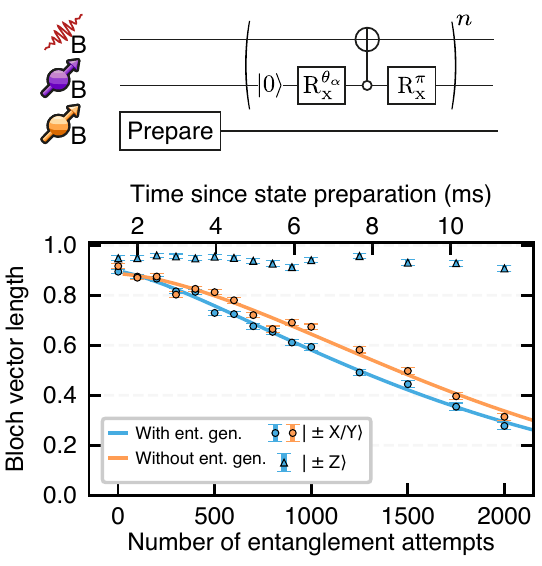}
\caption{\label{fig:memory}Memory qubit coherence under network activity. (Top) Circuit diagram displaying the experimental sequence, with n the number of entanglement attempts. (Bottom) Blue: measured decay of memory qubit eigenstates (triangles) and superposition states (circles) versus entanglement attempts, for $\alpha = 0.05$. Orange: measured superposition decay versus time in the absence of entanglement attempts. Solid lines are fits, yielding decay constants of $N_{1/e} = 1843 \pm 32$ $(2042 \pm 36)$ with (without) entanglement generation attempts (see Table~\ref{tab:supp_fit_results} and Supplementary Material for further details).}
\end{figure}

This higher field puts much stricter demands on the relative field stability in order to not affect the qubit frequencies; we achieve an order of magnitude reduction in field fluctuations by actively stabilizing the temperature of the sample holder, which in turn stabilizes the permanent magnet inside the cryostat (see Supplementary Material). Additionally, the higher magnetic field splits the two optical transitions used for electronic spin initialization, hindering fast qubit resets; the addition of a second initialization laser, frequency locked to the first one with an offset of \SI{480}{\mega\hertz}, enables us to maintain high-fidelity ($>0.99$) and fast (few microsecond) resets (see Supplementary Material).

We measure the fidelity of stored states on Bob’s memory qubit for a varying number of entanglement generation attempts (Fig.~\ref{fig:memory}). The two eigenstates ($\ket{\pm Z}$) do not show significant decay as we increase the number of entanglement generation attempts, as expected from the pure dephasing nature of the process~\cite{kalb_dephasing_2018}. The superposition states degrade with an average decay constant of $N_{1/e} \approx 1800$ attempts. To gain insight into the contribution of network activity to this decay, we repeat these measurements in the absence of entanglement attempts, in which case dephasing of the memory qubit is mainly due to uncontrolled interactions with nearby nuclear spins. We find this intrinsic dephasing time to be $T_2^* = \SI{11.6(2)}{ms}$, equivalent to the duration of $\approx 2000$ entanglement generation attempts. We conclude that the intrinsic dephasing accounts for most of the decay observed under network activity, indicating the desired robustness. For the experiments discussed below, we use a timeout of \num{450} attempts before the sequence is restarted, as a balance between optimizing entanglement generation rate and fidelity of the stored state. 

Executing protocols over quantum networks requires real-time feed-forward operations among the various nodes: measurement outcomes at the heralding station or at nodes need to be translated into quantum gates on other nodes. We implement an asynchronous bi-directional serial communication scheme between micro-controllers at the nodes, enabling both the required timing synchronization of the nodes as well as exchanging feed-forward information for the quantum network protocols (see Supplementary Material). Furthermore, we integrate the feed-forward operations with local dynamical decoupling protocols that actively protect the communication qubits from decoherence. The resulting methods enable us to run multi-node protocols in a heralded fashion: “flag” signals indicate in real time the successful execution of (sub)protocols and generation of desired states which are then available for further use, thus critically enhancing the efficiency and removing the need for any post-selection.

\begin{figure*}
\includegraphics{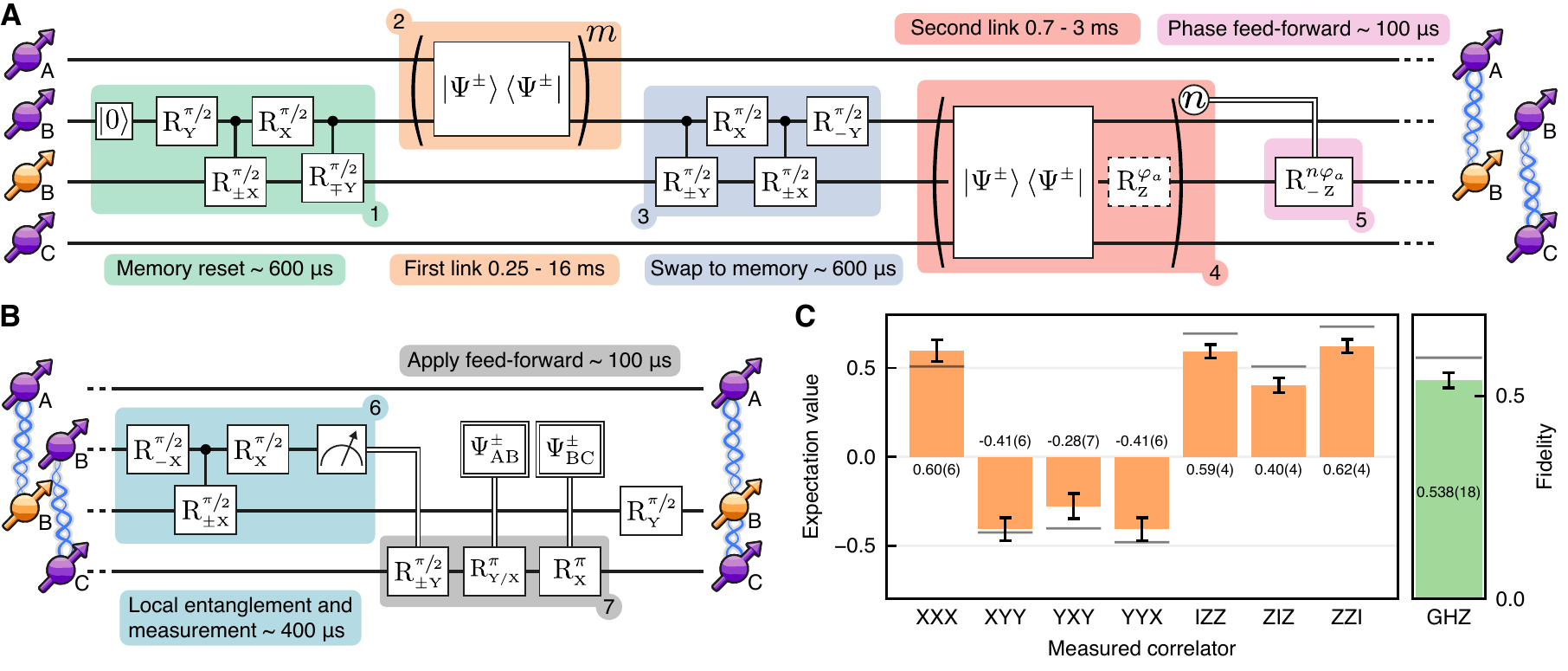}
\caption{\label{fig:ghz}Distribution of genuine multipartite entanglement across the quantum network. (A) Circuit diagram displaying the experimental sequence used to establish entanglement on both elementary links. (B) Circuit diagram displaying the experimental sequence for distributing a three-partite GHZ state across the three nodes. (C) Outcomes of correlation measurements and the resulting fidelity of the heralded GHZ state, demonstrating genuine multipartite entanglement.}
\end{figure*}

\section{Demonstration of multi-node network protocols}
We now turn to the full operation of the three-node network that combines the different elements discussed above. We perform two canonical network protocols: the distribution of genuine multipartite entanglement and entanglement swapping to two non-nearest-neighbor nodes. 

In both protocols, the sequence depicted in Figure~\ref{fig:ghz}A is used to establish a remote entangled state on each of the two links. This sequence starts with a preparation step (depicted only in Fig.~\ref{fig:supp_flowchart}) that synchronizes the micro-controllers of the nodes and makes sure that the NV centers in each node are in the desired charge state and in resonance with all the relevant lasers. After initialization of the memory qubit, the first entangled state is prepared on the link Alice-Bob. We interleave blocks of entanglement generation attempts with phase stabilization cycles. Once Alice-Bob entanglement is heralded, Alice’s entangled qubit is subject to a dynamical decoupling sequence while awaiting further communication from the other nodes. At Bob, deterministic quantum logic is used to swap the other half of the entangled state to the memory qubit. 

The second part of the phase stabilization is then executed, followed by the generation of remote entanglement between the communication qubits of Bob and Charlie. In case of a timeout (no success within the pre-set number of attempts), the full protocol is restarted. In case of success, a dynamical decoupling sequence is started on Charlie’s communication qubit analogous to the protocol on Alice. At Bob, a Z-rotation is applied to the memory qubit to compensate for the acquired phase that depends linearly on the (a priori unknown) number of entanglement attempts. This gate is implemented through a XY4 decoupling sequence on the communication qubit, with a length set in real time by the micro-controller based on which entanglement attempt was successful (see Supplementary Material). After this step, the two links each share an entangled state ready for further processing: one between the communication qubit at Alice and the memory qubit at Bob, and one between the communication qubits of Bob and Charlie.

\begin{figure*}
\includegraphics{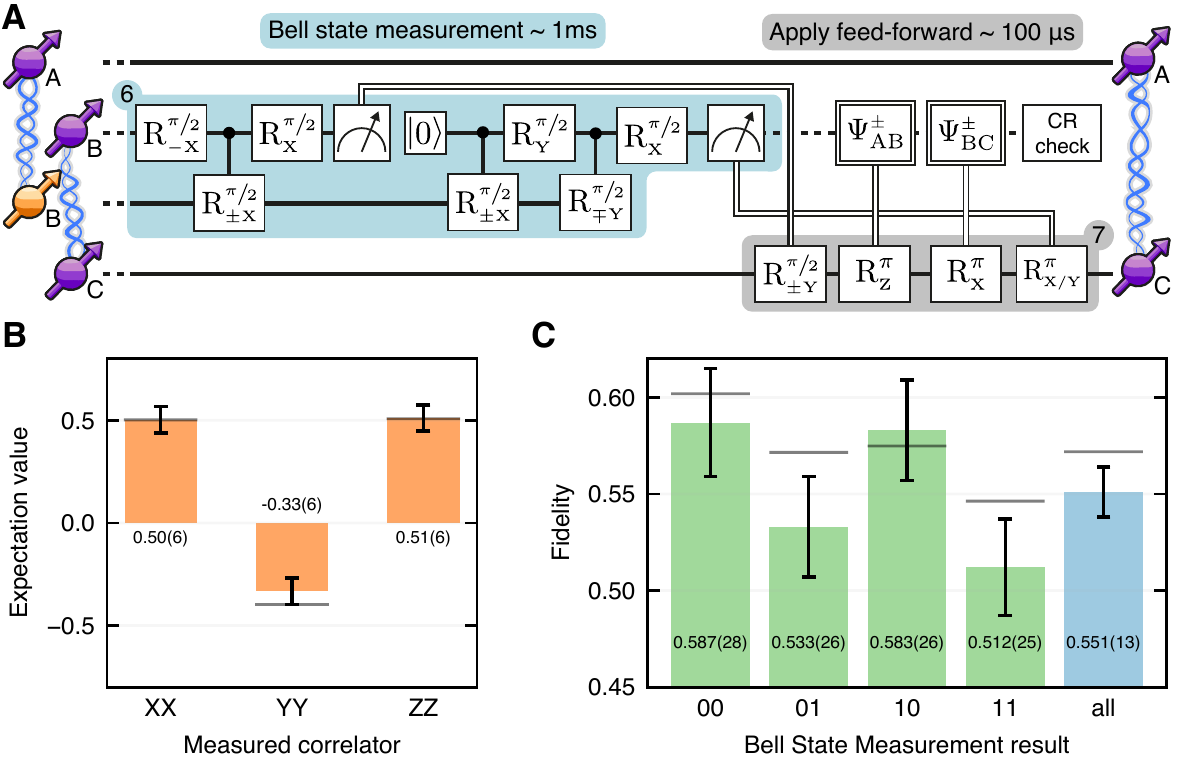}
\caption{\label{fig:swapping}Entanglement swapping on a multi-node quantum network. (A) Circuit diagram displaying the experimental sequence for entanglement swapping, yielding an entangled state shared between the two non-connected nodes. (B) Outcomes of correlation measurements on the heralded entangled state shared between Alice and Charlie for the selected Bell-state measurement outcome (see main text). (C) State fidelities for different outcomes of Bob’s Bell-state measurement (green) and the state fidelity averaged over all outcomes (blue).}
\end{figure*}

The first protocol we perform is the generation of a multipartite entangled GHZ state across the three nodes. The circuit diagram describing our protocol is depicted in Figure~\ref{fig:ghz}B. We first entangle the two qubits at Bob, followed by measurement of the communication qubit in a suitably chosen basis. The remaining three qubits are thereby projected into one of \num{4} possible GHZ-like states, which are all equivalent up to a basis rotation. The specific basis rotation depends both on the measurement outcome at Bob and on which Bell states ($\ket{\Psi^+}$ or  $\ket{\Psi^-}$) were generated in the first part of the sequence, which in turn depends on which two photon detectors heralded the remote entangled states. These outcomes are communicated and processed in real time and the corresponding feed-forward operations are applied at Charlie. As a result, the protocol is able to achieve delivery of the same GHZ state $\ket{\mathrm{GHZ}}_\mathrm{ABC} = \left( \ket{000} + \ket{111}\right)/\sqrt2$, irrespective of the intermediate outcomes. Here, we choose to herald only on Bob reporting the $\ket 0$ readout outcome, because the asymmetry in the communication qubit readout fidelities renders this outcome more faithful (see Supplementary Material). Additionally, this choice automatically filters out events in which the NV center of Bob was in the incorrect charge state or off resonance (occurrence $\approx \SI{10}{\percent}$ in this experiment, see Supplementary Material). With this heralding choice, the protocol delivers GHZ states at a rate of approximately $1/(\SI{90}{s})$.

We extract the fidelity to the ideal GHZ state from correlation measurements by using  $\mathrm{F} = ( 1 + \braket{IZZ} + \braket{ZIZ} + \braket{ZZI} + \braket{XXX} - \braket{XYY} - \braket{YXY} - \braket{YYX} ) / 8$  and find $\mathrm{F} = \num{0.538(18)}$ (Fig.~\ref{fig:ghz}C). The state fidelity above \num{0.5} certifies the presence of genuine multipartite entanglement distributed across the three nodes~\cite{guhne_entanglement_2009}. 

In this experiment, the fidelities of the entangled states on the elementary links bound the fidelity of the heralded GHZ state to about \num{0.66}. Other relevant error sources are dephasing of the memory qubit and accumulation of small quantum gate errors (see Table~\ref{tab:supp_budget_ghz}). We emphasize that, contrary to earlier demonstrations of distributed GHZ states with photonic qubits~\cite{bouwmeester_observation_1999} and ensemble-based memories~\cite{jing_entanglement_2019} that relied on post-selection, we achieve heralded GHZ state generation: a real-time heralding signal indicates the reliable delivery of the states.

The second protocol, illustrated in Figure~\ref{fig:swapping}A, demonstrates entanglement swapping of the two direct links into an entangled state of the outer two nodes. Once entanglement is established on the two links as described above, the central part of the entanglement swapping is executed: Bob, the central node, performs a Bell state measurement (BSM) on its two qubits. One way to read this protocol is that the BSM induces teleportation of the state stored on Bob’s memory qubit to Charlie, by consuming the entangled state shared by Bob’s communication qubit and Charlie. Since the state teleported to Charlie was Bob’s share of an entangled state with Alice, the teleportation establishes direct entanglement between Alice and Charlie. 

After the BSM is completed, we perform a charge and resonance (CR) check on Bob to prevent heralding on events in which the NV center of Bob was in the incorrect charge state or off resonance. We note that this CR check was not employed in the heralding procedure of the GHZ generation protocol because its current implementation induces decoherence on Bob’s memory qubit, which is part of the final GHZ state to be delivered.

To complete the entanglement swapping, feed-forward operations are performed at Charlie to account in real-time for the different measurement outcomes, analogous to the previous protocol, resulting in the delivery of the Bell state $\ket{\Phi^+}_\mathrm{AC} = \left( \ket{00} + \ket{11} \right) / \sqrt 2$. 

We assess the performance of the entanglement swapping by measuring three two-node correlators on the generated Bell state shared by Alice and Charlie. Since the BSM is performed with local quantum logic and single-shot readout, it is (except for the CR check step) a deterministic operation. However, given the asymmetry in the readout errors as discussed above, the fidelity of the final state will depend on the readout outcomes. Figure~\ref{fig:swapping}B shows the results of the correlation measurements on the delivered state for heralding on Bob obtaining twice the outcome $\ket{0}$, yielding a state fidelity of $\mathrm{F} = \num{0.587(28)}$. Figure~\ref{fig:swapping}C compares the state fidelities across the different BSM outcomes, displaying the expected lower fidelities for outcomes of $\ket 1$ and an average fidelity over all outcomes of $\mathrm{F} = \num{0.551(13)}$. The combined heralding rate is $1/(\SI{40}{s})$. The sources of infidelity are similar to the ones discussed above (see Table~\ref{tab:supp_budget_swapping}). This experiment constitutes the first demonstration of entanglement swapping from previously stored remote entangled states, enabled by the network’s ability to asynchronously establish heralded elementary entanglement links, to store these entangled states and then efficiently consume them to teleport entanglement to distant nodes.

\section{Conclusion}
In summary, we have demonstrated the realization of a multi-node quantum network. We achieved multipartite entanglement distribution across the three nodes and any-to-any connectivity through entanglement swapping. It is noteworthy that the data acquisition for the network protocols has been performed fully remotely due to the COVID19 pandemic, highlighting the versatility and stability of our architecture. 
Near-term advances in the capabilities and performance of the network will be driven by further reducing the infidelities of the elementary links, by adding new sub- protocols such as novel control methods~\cite{bradley_ten-qubit_2019}, decoupling sequences~\cite{kalb_dephasing_2018} and repetitive readout~\cite{jiang_repetitive_2009} for the nuclear spin qubits, by improved photonic interfaces to enhance the entangling rates~\cite{riedel_deterministic_2017, janitz_cavity_2020, ruf_resonant_2021}, and by improved control over the charge state of the NV center~\cite{baier_orbital_2020}.

Our results open the door to exploring advanced multi-node protocols and larger entangled states, for instance by extending the local registers at the nodes. We note that a fully controlled ten-qubit register has recently been demonstrated on a similar device~\cite{bradley_ten-qubit_2019}. Furthermore, the network provides a powerful platform for developing and testing higher-level quantum network control layers~\cite{van_meter_quantum_2014, pirker_quantum_2019, kozlowski_towards_2019}, such as the recently proposed link layer protocol for quantum networks~\cite{dahlberg_link_2019}. Finally, we expect the methods developed here to provide guidance for similar platforms reaching the same level of maturity in the future~\cite{rose_observation_2018, nguyen_quantum_2019, trusheim_transform-limited_2020, son_developing_2020}.

\begin{acknowledgments}
We thank Conor E.~Bradley, Sidney Cadot and Joris van Rantwijk for experimental support and Anders S.~Sørensen, Tracy E.~Northup, Johannes Borregaard and Tim H.~Taminiau for critically reviewing the manuscript.

We acknowledge financial support from the EU Flagship on Quantum Technologies through the project Quantum Internet Alliance, from the Netherlands Organisation for Scientific Research (NWO) through a VICI grant and the Zwaartekracht program Quantum Software Consortium, the European Research Council (ERC) through an ERC Starting Grant (S.W.) and a Consolidator Grant (R.H.). M.P.~acknowledges support from the Marie Skłodowska-Curie Actions - Nanoscale solid-state spin systems in emerging quantum technologies - Spin-NANO, Grant Agreement Number 676108. S.B.~acknowledges support from an Erwin-Sch\"odinger fellowship (QuantNet, No.~J 4229-N27) of the Austrian National Science Foundation (FWF).

The datasets that support this manuscript and the software to analyze them are available at \url{https://doi.org/10.4121/13600589}
\end{acknowledgments}

\bibliography{main}
\bibliographystyle{my_plain}
\onecolumngrid

\clearpage
\begin{center}
\textbf{\large Supplementary Materials}
\end{center}
\makeatletter
\renewcommand{\theequation}{S\arabic{equation}}
\renewcommand{\thefigure}{S\arabic{figure}}
\renewcommand{\thetable}{S\arabic{table}}
\renewcommand{\thesection}{S-\Roman{section}}
\makeatother

\setcounter{equation}{0}
\setcounter{figure}{0}
\setcounter{table}{0}
\setcounter{section}{0}
\section{Experimental setup}

Our experiments are performed on three quantum network nodes. Each node houses a Nitrogen-Vacancy (NV) center in a high-purity type-IIa chemical-vapor-deposition diamond cut along the $\langle 111 \rangle$  crystal orientation (Element Six).
All three samples have a natural abundance of carbon isotopes. 
Fabrication of solid immersion lenses and an anti-reflection coating on the diamond samples enhances the photon-collection efficiencies from the NV centers.
The samples are housed in home-built cryogenic confocal microscope setups at \SI{4}{K}. Experimental equipment used for each node is summarized in Table~\ref{tab:supp_equipment}.
Node A is in a different laboratory than Node B and C, \SI{7}{m} away. Node B and C are on the same optical table, approximately \SI{2}{m} apart, see also Figure~\ref{fig:network}A of the main text. The optical fiber that connects A with B is \SI{30}{m} long, while the one that connect B to C is \SI{2}{m} long.

\begin{figure}[ht]
    \centering
	\includegraphics{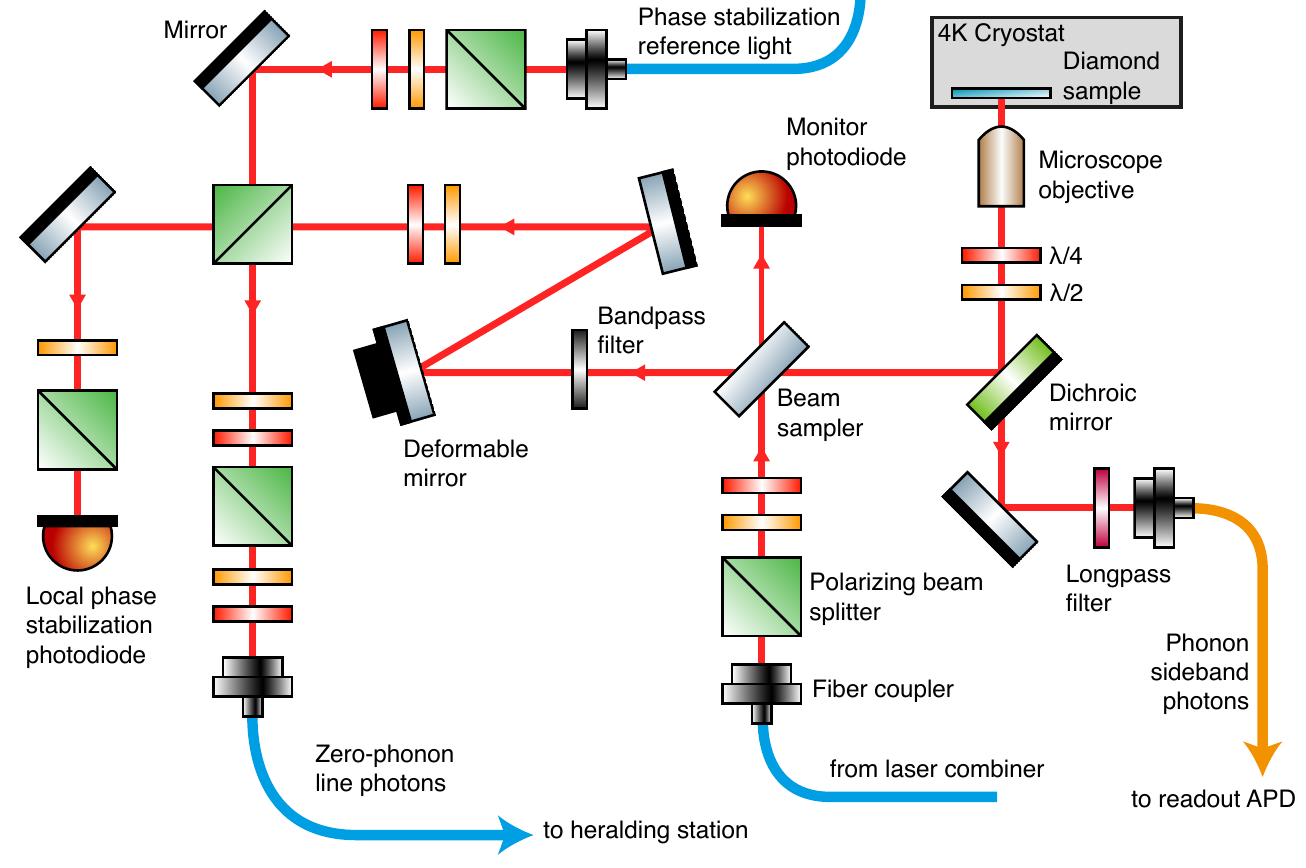}
	\caption{\label{fig:supp_optics}Schematic of the optics used for each node. The red lines indicate the optical path used both by the laser beams and the single photons. Blue fibers are single-mode polarization maintaining fiber. The orange fiber is a multi-mode fiber. The laser combiner (not depicted) combines, via beam-splitters and dichroic mirrors, the various laser beams and couples them into the single-mode fiber shown in the diagram. The laser combiner also includes a piezoelectric-mounted mirror that is used for the local phase stabilization feedback. The monitor photodiode records the \SI{90}{\percent} of excitation light that goes through the beam-sampler (and that would otherwise be discarded). We monitor this signal on a digital oscilloscope connected to the measurement computers for debugging purposes.
	}
\end{figure}

Figure~\ref{fig:supp_optics} depicts the optics used to deliver and collect light to each sample. For phonon-sideband (PSB) detection, a dichroic mirror (Semrock) and an additional long-pass filter (Semrock) are used to block reflections of the excitation lasers. Photon emission is detected via an avalanche photo-diode (APD, Laser components, quantum efficiency approximately \SI{80}{\percent}), with a total collection efficiency of approximately \SI{10}{\percent}. 
For zero-phonon line (ZPL) detection, we isolate the single photons first with a narrow bandpass filter (\SI{5}{nm}, Semrock), then by blocking the reflected excitation light via two polarising beam-splitters (Thorlabs and Semrock). Spatial mode shaping via a deformable mirror (Boston Micromachines) enhances coupling to a polarization-maintaining single-mode fiber.
The optical signals from each node are combined on in-fiber polarization-maintaining beam-splitters (Evanescent Optics). The final beam-splitter (where the single photons interfere) has an integrated fiber stretcher used for optical phase stabilization.
Finally, the single photons are detected on superconducting nanowire single photons detectors (Photon Spot). They are optimized for \SI{637}{nm}, have a detection efficiency $>$\SI{95}{\percent} and a dark count rate $<$\SI{1}{\hertz}.

\begin{figure}
    \centering
	\includegraphics[width=1\linewidth]{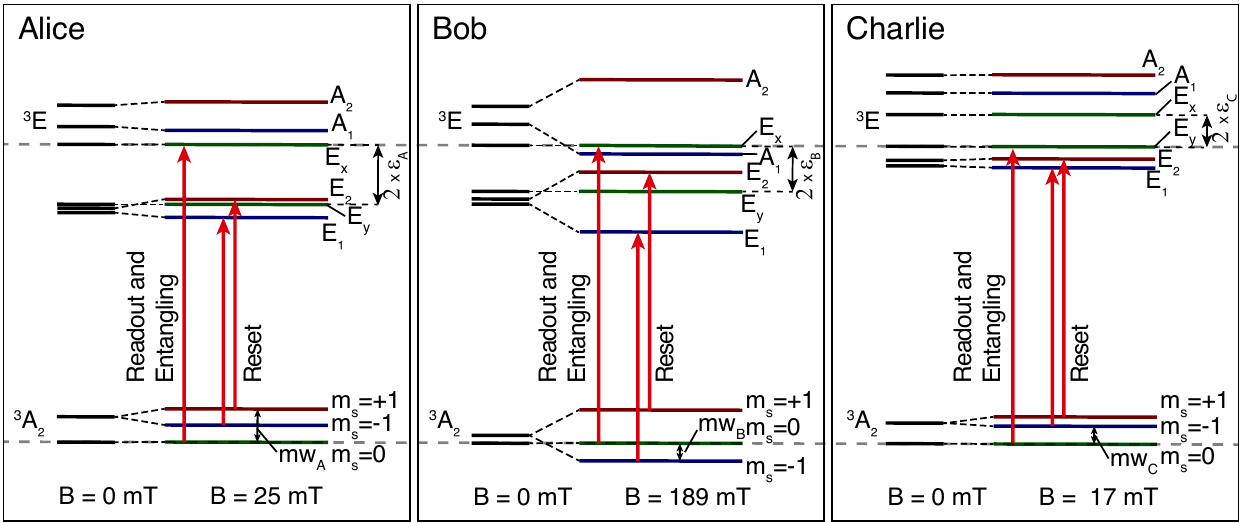}
	\caption{\label{fig:supp_levels}Level structure for the three NV centers. The optical transitions used within this work are indicated by red solid arrows. DC stark tuning brings all readout transitions to the same frequency, ensuring that the photons generated via the optical excitation pulse are indistinguishable. The spin state $m_S=0/+1/-1$ of each level is indicated by color (green/red/blue). The communication qubits of Nodes A, B and C are encoded in the NV center electronic spin states $\ket{0/1}_A \equiv \ket{m_S = 0/+1}$, $\ket{0/1}_B \equiv \ket{m_S = 0/-1}$ and $\ket{0/1}_C \equiv \ket{m_S = 0/-1}$. The memory qubit of Node B is encoded in the nuclear spin state of the addressed ${}^{13} C$ atom, $\ket{0/1} \equiv \ket{m_I = \pm \frac{1}{2}}$.
	}
\end{figure}

The level structures of the three nodes are depicted in Figure~\ref{fig:supp_levels}.
Each structure depends on local strain, electric fields and the applied magnetic field $B$.
For Node A and B the magnetic field is created with a permanent neodymium magnet inside the cryostat, which is located close to the sample and attached to the sample holder. The magnetic field is fine-tuned to be along the symmetry axis of the NV center using permanent neodymium magnets outside the cryostat. Node C has only a single permanent magnet outside the cryostat.

For optical excitation we set the laser frequencies (red arrows in Fig.~\ref{fig:supp_levels}) to the corresponding $^3A_2$ to $^3E$ transition.
Spin-selective excitation of ZPL transitions ($\lambda = \SI{637.25}{nm}$, $\omega=2\pi\times \SI{470.45}{\tera\hertz}$) enables qubit readout ("Entangling" in Fig.~\ref{fig:network}B of the main text, $m_{\rm s}= 0 \longleftrightarrow{} E_{\rm x/y}$) and qubit reset via optical spin-pumping ("Reset" in Fig.~\ref{fig:network}B of the main text, $m_{\rm s}= \pm 1 \longleftrightarrow{} E_{1,2}$).
While at low field a single laser is sufficient to address both qubit reset transitions, in case of Node B, which operates at \SI{189}{\milli\tesla}, we find a reset transitions splitting of \SI{480}{\mega\hertz}. An additional laser is implemented in order to drive both reset transitions efficiently.

In order to tune the readout transitions of each NV center into resonance we employ the DC Stark effect via DC-biasing the strip-line that is used to deliver microwave (MW) signals. The feedback sequence is analogous to the one used in Ref.~\cite{hensen_loophole-free_2015}. 
Node B operates at \SI{0}{V} tuning (it is grounded), and it uses non resonant charge reset with a green laser (\SI{515}{nm}). We observe small day to day drift in the readout frequency of Node B that we attribute to slow ice build-up on the sample as the transition frequency can be brought back to its original value by a warm-up cool-down cycle.
Setup A and C, which use resonant charge reset with a yellow laser (\SI{575}{nm}), are brought into resonance with Bob before starting a measurement.

The memory qubit of Bob is the nuclear spin of a ${}^{13}$C atom in the proximity of the NV center. Its electronic-spin-dependent precession frequencies are $\omega_0 = 2\pi \times$\SI{2025}{\kilo\hertz} and $\omega_{-1} = 2\pi \times$\SI{2056}{\kilo\hertz}, resulting in parallel hyperfine coupling of $A_\parallel \approx 2\pi \times$ \SI{30}{\kilo\hertz}. The nuclear spin is controlled using dynamical decoupling sequences \cite{kalb_entanglement_2017, bradley_ten-qubit_2019}. The conditional $\pi/2$-rotations on the nuclear spin are performed with \num{56} decoupling pulses with an inter-pulse delay of $2\tau = 2\times$\SI{2.818}{\micro\second}.

For synchronization purposes, the micro-controllers at each node (J\" ager ADwin-Pro II T12) share a common \SI{1}{\mega\hertz} clock.

To increase the on-off ratio of the AOM RF drivers, and therefore reduce unwanted light leakage, we use home-built fast (\SI{150}{ns} rise-time) RF switches, based on the HMC8038 (Analog Devices), to disconnect the RF drivers from the AOMs when no power should be delivered. 

We terminate the MW delivery line on each cryostat with a home-built MW envelope detector, that allows us to see on an oscilloscope the microwave pulses being delivered to each sample. We use this for debugging purposes.

Parts that are not mentioned in the description above are the same as in Refs.~\cite{hensen_loophole-free_2015, kalb_entanglement_2017, humphreys_deterministic_2018}.

\section{Model of the generated states}
\label{sec:supp_model}
The Python code to model all the generated states and to produce the figures in the main text can be found at  \url{https://doi.org/10.4121/13600589}. 
The Jupyter notebooks that generate the figures make direct use of that code to plot the simulated states. In the following we use the letters A, B and C to identify nodes Alice, Bob and Charlie. The numbers 0, 1 refer to the computational basis $\ket 0, \ket 1$. The communication qubits of Nodes A, B and C are encoded in the NV center electronic spin states $\ket{0/1}_A \equiv \ket{m_S = 0/+1}$, $\ket{0/1}_B \equiv \ket{m_S = 0/-1}$ and $\ket{0/1}_C \equiv \ket{m_S = 0/-1}$. The memory qubit of Node B is encoded in the nuclear spin state of the addressed ${}^{13} C$ atom, $\ket{0/1} \equiv \ket{m_I = \pm \frac{1}{2}}$.

Regarding the generation of Bell states on the Alice-Bob and Bob-Charlie links, we extend the model presented in Ref.~\cite{humphreys_deterministic_2018} to allow for different values of the parameters $\alpha$ in the two nodes.  We find that to obtain maximum state fidelity the condition $\alpha_A p^\mathrm{det}_A \approx \alpha_B p^\mathrm{det}_B$ must hold, where $\alpha_{A,B}$ are the populations of the $\ket 0$ state of each node and $p^\mathrm{det}_{A,B}$ is the probability of detecting a photon emitted by the respective node in the detection window. The state that is heralded by the protocol is the following (assuming $p^\mathrm{det}\ll 1$):
\begin{equation}
    \rho^\pm_{AB} = \frac{1}{p_\mathrm{tot}}\begin{pmatrix}
    p_{00} & 0 & 0 & 0\\
    0&p_{01}&\pm \sqrt{V p_{01} p_{10}}&0\\
    0&\pm \sqrt{V p_{01} p_{10}}&p_{10}&0\\
    0&0&0&p_{11}
    \end{pmatrix},
\end{equation}
\begin{align}
    p_{00} &= \alpha_A \alpha_B (p^\mathrm{det}_{A} + p^\mathrm{det}_{B} + 2 p_\mathrm{dc}), \\
    p_{01} &= \alpha_A (1 - \alpha_B) (p^\mathrm{det}_{A} + 2p_\mathrm{dc}),\\
    p_{10} &= \alpha_B (1 - \alpha_A) (p^\mathrm{det}_{B} + 2p_\mathrm{dc}),\\
    p_{11} &= 2 (1 - \alpha_A) (1 - \alpha_B) p_\mathrm{dc},\\
    p_\mathrm{tot} &= p_{00} + p_{01} + p_{10} + p_{11}
\end{align}
where $V$ is the visibility of the two-photon quantum interference, $p_\mathrm{dc} \ll 1$ is the probability of detecting a dark count (or in general a non-NV photon) in the detection window, the $\pm$ sign depends on which detector clicked. The off-diagonal terms neglect the contribution due to the dark counts with respect to the contribution due to $p^\mathrm{det}_{A,B}$, i.e. we assume $p_\mathrm{dc} \ll p^\mathrm{det}_{A,B}$. 

If one assumes $p_\mathrm{det} = p^\mathrm{det}_{A} = p^\mathrm{det}_{B}$, $p_\mathrm{dc} = 0$, $\alpha = \alpha_A = \alpha_B$ and $V = 1$, then the fidelity of $\rho_{AB}$ with the closest Bell state is $F = 1 - \alpha$, and the generation rate is $r_{AB} = 2~\alpha~p_\mathrm{det}$.

Additional sources of infidelity are uncertainty in the phase of the entangled state and double excitation. See Ref.~\cite{humphreys_deterministic_2018} for details on how they are modeled. We summarize in Table~\ref{tab:supp_budget_epr} the infidelity arising from the aforementioned sources, which is reasonably in agreement with the measured state fidelities. The Bell states between Alice and Bob were generated with $\alpha_A, \alpha_B = 0.07, 0.05$, while the ones between Bob and Charlie with $\alpha_B, \alpha_C = 0.05, 0.10$. These values have been chosen as a trade-off between protocol success rate and fidelity.

To model the states generated in the two protocols demonstration (GHZ state between Alice Bob and Charlie, and Bell state between Alice and Charlie) we take into account:\begin{itemize}
    \item the Bell states generated between Alice and Bob and between Bob and Charlie,
    \item the dephasing of the nuclear spin during the entanglement generation between Bob and Charlie,
    \item depolarising noise on the nuclear spin that combines initialisation, swap and readout error,
    \item communication qubit readout errors at Bob that would generate a wrong feed-forward operation at Charlie.
\end{itemize}
Table~\ref{tab:supp_budget_ghz} and Table~\ref{tab:supp_budget_swapping} summarize the error budget of the generated states. In the case of the entanglement swapping we also report the expected infidelity when accepting any Bell state measurement (BSM) result.

\section{Phase stabilization}
Inherent to an implementation where active phase stabilization is interleaved with free evolution time, there is a trade-off between phase stability (or fidelity of the entangled state) and the free evolution time (which is the time used for entanglement generation). The more often the system is stabilized, the higher the bandwidth of the stabilization and the lower the final uncertainty in $\Delta\theta$ will be. \\
Our previous implementation used a single homodyne phase detection scheme \cite{humphreys_deterministic_2018}. While that method allows for the stabilization of the phase of the entangled state, there are several aspects that can be improved; The small fraction of excitation light that is reflected from the diamond surface is partially coupled in the single-photon detection path. By measuring the interference signal after the beam-splitter at the heralding station it is possible to obtain the phase of the interferometer. Leaking some of the reflected excitation light into the single-photon path for phase stabilization purposes increases the chance that, during entanglement generation, some of the reflected excitation light will be detected and mistakenly herald an entangled state.
To counteract this effect, the amount of leaked light was somewhat minimized by polarization selection (but never completely, since some light is needed to detect a phase signal) and long integration times (\SI{24}{ms}) were used during phase detection, reducing the phase stabilization bandwidth.
Furthermore, exposing the NV center to a relatively long and strong laser pulse makes it more susceptible to spectral jumps and ionization. 

To solve these challenges we devised and implemented a new phase stabilization scheme that combines higher bandwith and optimal rejection of the excitation light from the single-photon paths, while maintaining robustness against power level fluctuations and scalability to a higher number of nodes.
	
\subsection{Phase detection methods}
\subsubsection{Homodyne phase detection}\label{sec:supp_homodyne}
In a homodyne phase detection scheme the light has the same frequency in both arms of the interferometer. Depending on the optical phase difference  $\Delta\theta$, light will constructively or destructively interfere on the output ports of the beam-splitter. Assuming common polarization and perfectly overlapping spatial modes, the intensity $I_{3,4}$ in the output ports is
\begin{equation}
I_{3,4} = I_1 + I_2\pm2\sqrt{I_1I_2}\cos\Delta\theta.
\end{equation}
For known input levels $I_{1,2}$, $\Delta\theta$ can be calculated from the difference in intensity in the output ports of the beams-splitter. Fluctuations in the intensity of the input signals will lead to an error in the phase measurement, except for the case $\cos \Delta\theta = 0$ which gives $I_3 = I_4$ independent of the input intensity.

\subsubsection{Heterodyne phase detection}\label{sec:supp_heterodyne}
In a heterodyne phase detection scheme the light has different frequencies in the two arms of the interferometer. Again, assuming common polarization and perfectly overlapping modes, the light will interfere in the output ports resulting in a signal with amplitude
\begin{equation}
I_{3,4} = I_1 + I_2\pm\sqrt{I_1I_2}\big(\cos((\omega_1-\omega_2)t - \Delta\theta ) + \cos((\omega_1+\omega_2)t + \Delta\theta )\big)
\label{eq:het}
\end{equation}
where $\omega_{1,2}$ are the angular frequencies of the light. 
When we pick a relatively small frequency difference, $(\omega_1-\omega_2)/2\pi \approx $ \SI{10}{\mega\hertz}, we can ignore the last term in Eq. \ref{eq:het} and the resulting \SI{10}{\mega\hertz} beat signal can be measured with a photodiode and efficiently filtered from the DC background signal (the last term of Eq. \ref{eq:het} will have a frequency in the optical domain and will not be picked up by the photodiode due to the limited bandwidth). The phase of this beat signal corresponds to the optical phase difference in the two paths. Since the phase information is not translated to the amplitude of the beat signal, fluctuations in the input intensity  will not cause an error in the measurement. Moreover, this method is very suitable to measure small signals: if the signal is very small in one of the arms, the amplitude of the beat signal can be increased by increasing the intensity in the other arm.
 
 \subsection{Splitting the interferometer in parts}
In the experiments with three quantum nodes we have two effective interferometers that share part of their optical paths. We split the interferometers into six parts, see Figures \ref{fig:supp_phase_diagram_total}, \ref{fig:supp_phase_diagram_sub}. In total there are four local interferometers and two global interferometers, where the local interferometer comprises the excitation path and free space optical path close to the cryostat of each node and the global interferometer includes the fibers connecting the nodes to the central beam-splitter. With the measured phase, an error signal is computed and feedback is applied to the optical path, either with a mirror on a piezoelectric element or a fiber stretcher.

The two global interferometers, using homodyne phase detection, stabilize the optical path to the beam-splitter and single photon detectors used for entanglement heralding. Since the detectors are shared for the two entanglement links, the optical phase measurement for the two global interferometers has to be multiplexed in time.
The local interferometers are stabilized using heterodyne phase detection. The excitation light (the same we use for the optical excitation pulse that generates spin-photon entanglement) is reflected off the diamond surface and since it has (close to) orthogonal polarization with the NV centers emitted photons it can be separated from the single photons using a polarizing beam-splitter (PBS). Afterwards, the weak reflected excitation pulse interferes with a strong laser pulse from the other arm with the frequency offset. The beat signal is measured with a photodiode and the optical phase difference is extracted using an electronic reference signal. The middle node has two local interferometers, one for each link.
When all separate interferometers are stabilized, the paths of the excitation light and the single photons used for entanglement heralding will be phase stable.

\subsection{Technical description of the local interferometer}
 For all the local interferometers we use a heterodyne phase detection scheme. A diagram of the optics and electronics is plotted in Figure~\ref{fig:supp_phase_two_node}. For each entanglement link (Alice-Bob and Bob-Charlie) the phase and excitation light are provided by the outer nodes (Alice and Charlie). 
 To generate the known \SI{10}{\mega\hertz} frequency offset between the light paths, we take advantage of the acousto-optic modulators (AOMs) we use to generate light pulses. By driving two AOMs at respectively \SI{200}{\mega\hertz} and \SI{210}{\mega\hertz}, we establish the required frequency difference between the light paths. Part of the RF signals used to drive the AOMs are tapped off and combined in a mixer to obtain an electronic reference signal. The light from the AOMs in launched in a free space path with several optical elements. The first polarizing beam-splitter (PBS) ensures the phase light to be linearly polarized. The second PBS separates the reflected excitation light from the single photons. At this point the phase-reference light and the reflected excitation light have orthogonal polarization. The waveplates in front of the third beam-splitter rotate their polarization such that they can be interfered on the third PBS. This interference leads to a beating signal that can be detected with the photodiode. Consequently, the beating signal is filtered, amplified and, together with the electronic reference signal, used as inputs for the phase detector (Mini-Circuits ZRDP-1+). The output of the phase detector is filtered and impedance matched to an analog to digital converter (ADC) input of the micro-controller, the ADwin.
 
\subsection{Timings}\label{sec:supp_timings}
 The phase stabilization requires synchronization between the different nodes. Node A and C provide the phase and excitation light, but all nodes measure the phase of at least one local interferometer. Some of the detectors used for the phase measurements are shared among different interferometers, so not all measurements can be done at the same time. Figure \ref{fig:supp_phase_timings} shows how the various phase stabilization cycles are interleaved with entanglement generation time.
 
 The choice of free evolution time is governed by the noise sources in the different parts of the system. The local interferometer of Node A experiences noise with high frequency components (compared to the other intereferometers) hence the free running time must be short enough to achieve the necessary feedback bandwith. The duration of the preparation part of the experiment, which includes charge and resonance checks, as well as synchronization steps between nodes, can vary from approximately \SI{50}{\micro s} to a few seconds. When the phase is completely scrambled due to a too long free running time, it is not possible to reach the set-point in a single feedback round. For this reason we start with multiple  rounds of phase stabilization without any free evolution time in between. 

\subsection{Phase stability}
To characterize the performance of the phase stabilization we look at three different aspects: the free evolution of the phase without any stabilization, the frequency spectrum of the noise and the distribution of the phase while actively stabilizing.
All the results for the six interferometers are plotted in Figure~\ref{fig:supp_phase_results} (see Fig.~\ref{fig:supp_phase_diagram_sub} for the labeling). The differences in performance can be explained by the noise sources present in our experimental lay-out. We identify two main sources of noise: the relatively noisy fiber connection between  Nodes A and B and the positioning stages of each node. The three nodes are built in two separates rooms and we use optical fibers (\SI{30}{m}) to connect Node A to Node B. All nodes have a microscope objective for optically accessing the diamond samples. On Node B and C this microscope objective is mounted on a piezo-electric stage. For Node A the design is different: here the sample is mounted on a piezo-stack and the microscope objective is fixed. All these piezo-electric stages are susceptible to the vibrations generated by the operation of the cryostats.

The sample stage of Node A cause relatively-strong high-frequency ($>$\SI{500}{\hertz}) noise; the microscope objective stage of Node B and C cause lower-frequency noise and the optical fiber connection between Node A and B causes relatively-strong low-frequency components. In the experimental sequence we interleave experimental time with rounds of phase stabilization. With the used timings (see Fig.~\ref{fig:supp_phase_timings}) we are able to stabilize frequencies $\leq$ \SI{500}{Hz}. Due to its relatively high-frequency components, the noise of the local interferometer of Node A is the limiting factor in terms of phase stability of the overall apparatus.
We expect that fixing the sample to the cold-finger of the cryostat, and only moving the microscope objective (like we do on Node B and C) will allow us to lower the phase noise on Node A in the future.  

\subsection{Entangled-state phase drifts}
While the phase stabilization scheme allows us to access the entangled state generated by the single photon protocol by fixing the phase $\Delta\theta$, we observe that the phase of the generated entangled state undergoes small drifts on a  timescale of hours. That is, even though all the interferometers are stabilized to the same value, the phase of the entangled state will slowly drift by $\approx$\SI{10}{\deg/\hour}.
We hypothesize that these drifts are due to the relative position of the microscope objective and the NV center: while the light used for phase stabilization is reflected off the diamond surface, the NV-emitted photons are generated inside the diamond. Small changes in distance and angle of the microscope objective would not lead to observable differences in the fluorescence measurement we use for position optimization, but may slightly alter the path the photons have to travel. To solve this challenge, after every position optimization ($\approx$ once every hour), we re-calibrate the phase of the generated entangled state ($\approx$ 5 minute measurement per link). More robust positioning systems (both for the sample and the microscope objective) may reduce the phase drifts and alleviate the need for entangled-phase re-calibration

\section{Single-shot readout correction}
We correct tomography-related single-shot readouts (SSROs) for known error in order to obtain a reliable estimate of the actual generated states.
\subsection{Single setup case}
For a single qubit:
\begin{equation}
\label{eq:single_qubit_exp}
\mathbf M = \hat R \ \mathbf P,
\end{equation}
where $\mathbf P = (p_0, p_1)^T$ is the (column) vector of expected populations, $\mathbf M = (m_0, m_1)^T$ is the (column) vector of measured populations, and 
$$
\hat R =  \begin{pmatrix}
r_{00} & r_{01}  \\
r_{10} & r_{11}
\end{pmatrix} = 
\begin{pmatrix}
F_0 & 1 - F_1  \\
1 - F_0 & F_1
\end{pmatrix}
$$
is the SSRO operator that connects the two. For example:
$$
m_0 = F_0 p_0 + (1-F_1) p_1,
$$
i.e. the measured population in $\ket 0$ is given by the correctly assigned population in $\ket 0$ plus the incorrectly assigned population in $\ket 1$. From \autoref{eq:single_qubit_exp} it follows that:
\begin{equation}
\label{single_qubit_inv}
\mathbf P = \hat R^{-1} \mathbf M,
\end{equation}
which is what we use in practice to apply the readout correction. This allow us to obtain the vector of expected populations given the measured populations and the SSRO error operator. 
Experimentally we cannot directly measure $\mathbf{M}$. We measure events in which the communication qubit is either in $\ket 0$ or in $\ket 1$. We repeat this process $N$ times, obtaining $N_0$ times the outcome $\ket 0$ and $N_1$ times the outcome $\ket 1$.
From this we estimate the measured populations $\mathbf M$:
\begin{equation}
m_0 = N_0 / N, \\
m_1 = N_1 / N
\end{equation}
The probability distribution of the number of events $N_0$ is a Binomial distribution with expected value $N m_0$ and variance $N m_0 (1-m_0)$.
From this it is possible to calculate the experimental value and uncertainty for $m_0$ (and $m_1$):
\begin{align}
    m_0 &= N_0 / N \\
    m_1 & = 1 - m_0 \\
    \sigma_{m_0} = \sigma_{m_1} &= \sqrt{\frac{m_0}{N} \left( 1- m_0 \right)}
\end{align}
The covariance between $m_0$ and $m_1$ is:
\begin{equation}
    \text{Cov}(m_0 ,m_1) = -\frac{m_0 (1 - m_0)}{N}
\end{equation}
Once \autoref{single_qubit_inv} has been calculated it is possible to evaluate the expectation value of $\mathbf{P}$ and its uncertainty.
In the one qubit scenario it is easy to invert the expression analytically:
\begin{align}
    p_0 &= \frac{F_1\ m_0 + (F_1 - 1)\ m_1}{F_0 + F_1 - 1} = \frac{F_1 + m_0 - 1}{F_0 + F_1 - 1}\\ 
    p_1 &= \frac{(F_0 - 1)\ m_0 + F_0\ m_1}{F_0 + F_1 - 1} = \frac{F_0 - m_0}{F_0 + F_1 - 1} = 1 - p_0\\
    \sigma_{p_0} = \sigma_{p_1} &= \frac{\sigma_{m_0}}{F_0 + F_1 - 1}
\end{align}
and it is straightforward to propagate uncertainties in $F_{0/1}$ to $p_0$ and $p_1$.
\subsection{Two and three setup case}
For two (and more) qubits, the measurement outcomes will be distributed according to a Multinomial distribution (as opposed to a Binomial). While the expectation values of $p_0, \ldots, p_i$ can still be computed analytically relatively straightforwardly, their uncertainties need to take into account the non-trivial covariances in the $m_i$. Additionally, taking into account uncertainties in the $F_{0/1}$ makes the error propagation even more tedious. We therefore use a Monte Carlo simulation that takes into account the Multinomial distribution as well as the $F_{0/1}$ of each setup to estimate uncertainties on the correlation measurements and the state fidelities, without having to assume normality of the data. The code to run the Monte Carlo simulation is included in the Jupyter notebooks that produce the figures of the main text.

\section{Phase feed-forward on the nuclear spin}
The nuclear spin memory qubit of Bob precesses at a frequency that depends on the spin state of the electronic spin (the communication qubit). Throughout the experimental sequence we keep track of the phase acquired by the nuclear spin to be able to readout and apply gates in the correct bases. While most operations are deterministic in time (nuclear spin initialisation, gates on the electronic spin, etc.) and the phase evolution of the nuclear spin can be calculated in advance, entanglement generation is a probabilistic process. This means that it is not known in advance how long the entanglement operation (number of entanglement attempts) is going to take, and therefore how much phase the nuclear spin is going to acquire. 
To solve this challenge, we implement a phase feed-forward mechanism that applies a Z-rotation to the nuclear spin that cancels this acquired phase. Since the used Arbitrary Waveform Generator (AWG) only has limited real-time programming capability, we implement this mechanism via a real-time interaction between our node micro-controller (ADwin) and the AWG. 
Once entanglement is heralded between Bob and Charlie, the AWG of Bob jumps out of the entanglement generation subroutine and starts an XY4 decoupling sequence on the communication qubit. During this XY4, the AWG interacts with the ADwin of Bob (which has recorded how much phase the nuclear spin has acquired during the entanglement operation) to select, via a binary decision tree, the time in between microwave pulses. The binary decision tree allows us to vary the (additional) duration of the XY4 element in steps of \SI{2}{ns} up to \SI{512}{ns}, which is more than a $2\pi$ precession for the nuclear spin ($\tau_L = $\SI{490}{ns}, feed-forward resolution $\approx$ \ang{1.5}). Regardless of the inter-pulse time selected, the communication qubit will be decoupled. Consequently, the needed additional phase to re-phase the nuclear spin can be conveniently set via the length of the XY4 sequence. 
We pre-compile the timings that the ADwin will communicate to the AWG to reduce the computational load on the ADwin. We anticipate that an AWG with an integrated programmable FPGA will be able to completely take over the task of phase tracking without need for interaction with the node micro-controller, reducing experimental overhead.

\section{Memory lifetime and high magnetic field}
\label{sec:supp_memory_lifetime}
An important resource in our experiments is the ability to store entanglement in the nuclear spin memory qubit of Bob while performing further operations on the node.
While we have implemented methods to keep track and actively compensate for the phase acquired during entanglement generation (see previous section), additional dephasing may occur. The major source of nuclear spin dephasing during entanglement generation was found to be \cite{kalb_dephasing_2018} failed electronic spin control (initialization errors or MW pulse errors).

An entanglement attempt can be broken into the following pieces: communication qubit reset (via optical pumping), MW pulse that creates the communication qubit superposition (named in the following the $\alpha$ pulse), optical excitation pulse that creates the spin-photon entanglement, and a decoupling MW $\pi$ pulse.
The time $\tau$ between the $\alpha$ and the de-coupling pulse is chosen such that it equals the time between the decoupling pulse and the average reset time in the subsequent entanglement attempt (see Ref.~\cite{kalb_dephasing_2018} for details). This ensures that regardless of its initial state, the communication qubit spends an equal amount of time in the $\ket{0}$ and $\ket{1}$ states.
However, an error in the MW $\pi$ pulse will result in an unknown acquired phase on the nuclear spin and lead to dephasing.
Previous work \cite{kalb_dephasing_2018} suggested that such dephasing can be mitigated when working at a higher magnetic field, which allows for a shorter spacing between subsequent MW pulses.

In order to work at higher fields we have installed a stronger permanent magnet inside the cryostat of Bob reaching a field of \SI{189}{\milli\tesla} at the location of the NV center. 
At such fields, temperature fluctuations of the magnet, mainly due to the MW pulses applied to the sample, can result in a significant change of the magnetic field amplitude. Hence, we stabilize the sample holder via an active feedback loop, ensuring a stable temperature of the permanent magnet. We reach a stability of \SI{1}{\micro\tesla}, which results in a maximum variation of the nuclear spin precession frequency of $\approx$ \SI{10}{\hertz}, one order of magnitude below the dephasing rate due to interactions with other spins in its environment.

These improvements allow us to shorten the interpulse spacing to \SI{942}{\nano\second}, limited by the waiting time after the optical excitation pulse that we need to include in order to allow the AWG to respond in real time to a successful entanglement attempt and jump out of the entangling sequence.
As Figure~\ref{fig:memory} of the main text shows, a similar nuclear memory lifetime is observed when applying entanglement attempts or when idling. This shows that the lifetime of the memory qubit, in our magnetic field regime, is mainly limited by natural dephasing and not by electronic spin control errors. We fit the two decays with the following function:
\begin{equation}
    f(N) = A \exp\left(-\left(\frac{N}{N_{1/e}}\right)^n\right),
\end{equation}
with $N$ the number of entanglement generation attempts, $N_{1/e}$ the $N$ at which the Bloch vector length has decayed to $1/e$ of its initial value $A$, and $n$ the exponent of the decay. The results of the fit are reported in Table~\ref{tab:supp_fit_results}. For the results \textit{Without ent. gen.} the entanglement generation attempt is replaced by the equivalent free evolution time.

\section{Microwave pulse fidelity}
During entanglement generation, errors in the MW pulses induce decoherence on the nuclear spin memory qubit \cite{kalb_dephasing_2018}. 
We use Hermite MW pulse envelopes \cite{bradley_ten-qubit_2019} to perform rotations of the communication qubit spin:
\begin{equation}
	h(t) = \left( 1 - p \left(\frac{t}{T}\right)^2 \right) e^{- \left(\frac{t}{T}\right)^2},
\end{equation}
where $p$ affects the shape of the pulse and $T$ changes the length of the pulse. The pulses get distorted by the transmission line before they get to the sample. We apply a linear pre-distortion in frequency domain to compensate part of the error via the following IQ signals:
\begin{align}
	I &= a \cdot h(t)\\
	Q &= a b \frac{t}{\pi T^2} \left( p + 1 - p \left(\frac{t}{T}\right)^2 \right) e^{- \left(\frac{t}{T}\right)^2},
	\label{eq: i-and-q-channel}
\end{align}
where $a$ is the amplitude of the pulse and $b$ is the skewness (slope) of the pre-distortion in frequency domain.

\begin{figure}
	\centering
	\includegraphics[width=.6\textwidth]{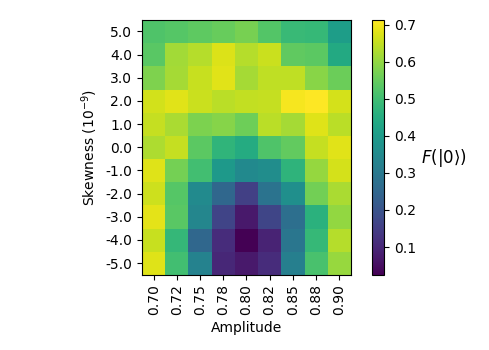}
	\caption{\label{fig:supp_skewness}Calibrating the pre-distorted microwave (MW) Hermite pulses. The $\pi$ pulses are calibrated by applying 11 sequential pulses: the probability of being in $\ket 0$ at the end of the sequence is measured for different amplitudes ($a$) and skewness ($b$) of the Hermite pulse. The linear frequency pre-distortion allows us to achieve lower errors for the MW pulses.}
\end{figure}

The MW $\pi$-pulses are calibrated by initializing the qubit in the $\ket 0$ state, applying an odd number of consecutive pulses and reading out the final state. If the pulses were perfect one would measure $\ket 1$ as outcome. The effect of the skewness on the pulse fidelity is investigated with a two dimensional scan; evaluating the fidelity for pulses with different amplitudes ($a$) and skewness ($b$). 
Figure~\ref{fig:supp_skewness} shows an example of such a scan, where it is clear one can calibrate $a$ and $b$ almost independently.
We find that different set-ups require different levels of pre-distortion $b$, ranging from \num{e-11} to \num{e-8}. We estimate that the current errors of our MW pulses are between \SI{0.1}{\percent} and \SI{1}{\percent} for all the three nodes.

\section{Classical communication}
The three nodes can share information in several ways. The slowest method is based on Python socket interfaces between the measurement computers that allow us to share necessary values and information at a rate of approximately \SI{10}{\hertz}; this method is used for example to frequency lock the lasers, to coordinate calibrations on all nodes from a single computer and to share and record environmental data such as the temperature in the different laboratories.
The second, and fastest, method is a direct connection between the micro-controller and the AWGs. This enables the triggering of all the AWGs from a single node, reducing jitter on the output waveforms.
The third method is implemented on the micro-controllers and is used for the feed-forward operations across the nodes. Each micro-controller has one input and one output communication port (physically it is a normal digital input-output coaxial port). Bob, which receives signals from both Alice and Charlie, has a digital summing box (OR gate) at its input port, that combines the signals coming from the other two nodes. We designed the experimental sequence such that it is clear who sent a specific message depending on when it arrives.
Messages are sent over an off-the-shelf coaxial cable, using a serial communication scheme, with an average bit interval of \SI{60}{ns} (the shortest the micro-controller can achieve). At the input port of each micro-controller, a fast edge detection (\SI{100}{\mega\hertz}) stores changes in the signal level (and the time at which they occur). It is therefore possible to reconstruct what pattern (i.e. message) was sent from one node to the other, directly on the micro-controller. Sending a message takes up to \SI{300}{ns} (we send up to 5 bits at a time). Receiving and decoding take up to \SI{2}{\micro\second} combined.

\section{Feed-forward operations between nodes}
We implement the feed-forward operations needed for our experimental protocols by combining the classical communication just discussed with a real-time pulse selection sequence by the micro-controller on the AWG. For both network protocols demonstrated we need to apply gates on the communication qubit of Charlie conditional on measurement outcomes at Bob. 
Once Bob has performed the required readout operations (on the communication qubit for the GHZ state generation or on both qubits for the entanglement swapping) it combines the readout results with the Bell states generation outcomes (i.e. which detectors clicked in the A-B and B-C entanglement generation) to obtain one of four possible feed-forward messages. 
Combining this information on Bob is an optimization of our communication resources; we could, alternatively, send the bits of information one by one to Charlie and combine the information there should that be a requirement of the protocol (for example in a blind quantum computation scenario). At this point a \textit{FAIL} message could also be sent from Bob to all the nodes in order to abort the whole sequence, for example if the Bell State Measurement result is not the one that gives high-fidelity (see main text). We choose to not send \textit{FAIL} messages and instead continue with the protocol to be able to assess the protocol performance for the less faithful Bell State Measurement outcomes (see Fig.~\ref{fig:swapping}C of the main text). 
In the meantime, Charlie has been applying an XY8 decoupling sequence to the communication qubit to protect its coherence while Bob performed the readout operations. Once Charlie receives the feed-forward information, its micro-controller starts a decision-tree sequence with its AWG to select the required microwave pulse-sequence. This decision tree is incorporated into an XY8 block of the AWG, such that the slow response time of the AWG (\SI{1}{\micro\second} per bit of information) does not affect the coherence of the communication qubit. The microwave pulse sequence selected via the decision tree is appended to the aforementioned XY8 block. 
After the feed-forward operations are performed, the delivery of the states by the network protocol is completed. Finally, the delivered states are analyzed using a readout sequence (composed of an optional basis rotation and state readout).

\section{Data acquisition and calibrations}
The data supporting the protocol demonstrations in the main text (Figures~\ref{fig:ghz}C,~\ref{fig:swapping}B,~\ref{fig:swapping}C) was gathered in the month of October 2020. Due to the restrictions imposed by the COVID19 pandemic, we operated the setups remotely (from home) and went to the laboratories only when something needed in-situ intervention (like a broken power-supply).

The data has been collected in blocks of approximately \SI{1}{\hour}, interleaved by calibration routines of approximately \SI{20}{\minute}.
For the GHZ state generation protocol we set the target number of data points at \num{2000}. For the entanglement swapping we set the target number of data points at \num{4000}. We stopped the experiment once the measurement block was completed in which the target number of data points was surpassed.

For the GHZ state generation we acquired \num{55} blocks over \num{10} days (effective measurement time $\approx$ \SI{50}{\hour}), obtaining \num{2028} events, equivalent to a rate of $r_{\mathrm{GHZ}} \approx (\SI{90}{\second})^{-1}$.

For the entanglement swapping demonstration we acquired \num{53} blocks over \num{7} days (effective measurement time $\approx$ \SI{45}{\hour}), obtaining \num{853} events with BSM result ``00", equivalent to a rate of $r_{\mathrm{swapping}} \approx (\SI{3}{minutes})^{-1}$. The other BSM results were: ``01'': \num{1030} events, ``10'': \num{1004} events, ``11'': \num{1168} events. The ratio of events between the BSM results matches the readout characteristics of Node B: measured (expected) share of the events, $0.21:0.25:0.25:0.29$ ($0.23:0.25:0.25:0.27$).
Combining all the BSM results we obtained a total of \num{4055} events, equivalent to a rate $r'_{\mathrm{swapping}} \approx (\SI{40}{\second})^{-1}$.

Every three measurement blocks we performed a fidelity check on the entangled states between Alice-Bob and Bob-Charlie at the target $\alpha$ (total duration \SI{20}{\minute}). These fidelity checks, combined over the GHZ and Entanglement Swapping datasets, are used for Figure 2E of the main text. We performed a total of 58 fidelity checks, that combined generated: \num{24197} $\Psi^+_\mathrm{AB}$ events, \num{25057} $\Psi^-_\mathrm{AB}$ events, \num{26383} $\Psi^+_\mathrm{BC}$ events and \num{27459} $\Psi^-_\mathrm{BC}$ events.
The asymmetry in the number of events between the $\Psi^+$ and the $\Psi^-$ states is due in part to the beam-splitter having a non ideal splitting ratio ($0.493 : 0.507$), to a slight difference in detector efficiencies ($\approx \SI{1}{\percent}$) and to the brightnesses ($\alpha~p_\mathrm{det}$) of the two setups involved not being completely balanced. The asymmetry between the number of events for the $\Psi_\mathrm{AB}$ and the $\Psi_\mathrm{BC}$ states is due to the different probability for Node B to be in the wrong charge state (NV${}^0$) at the end of the sequence for the two links. To obtain a reliable estimate of the fidelities of the Bell states, we discard events in which a CR (charge and resonance) check performed after readout gives a negative result.
We remark that, as mentioned in the main text, we do not perform such an operation for the network protocols demonstrations, which are free from any post-selection.
For the GHZ state generation, by heralding only on the $\ket 0$ readout outcome of the communication qubit of Bob, we automatically reject events in which the NV center of Bob was either in the wrong charge state or off resonant.
For the Entanglement Swapping demonstration, we perform a CR check after the Bell state measurement is performed on Node B, and we herald success of the whole protocol only if this final CR check gives a positive result. We find that the test gives a positive result in approximately \SI{90}{\percent} of the cases.

\section{Experimental monitoring} 
Analogous to what reported in section J of the SI of Ref.~\cite{hensen_loophole-free_2015}, we implement checks while the experiment is running to ensure that the nodes are performing as expected. If one of the checks does not pass, we mark all future data to be disregarded (until the check is passed) and / or pause the experiment to perform further calibrations. Following is a list of all the checks that we use to mark future data to be disregarded (if they don't pass):
\begin{itemize}
    \item Check that the measured phase of each interferometer is below \ang{50} before the last piezo feedback is performed.
    \item Check that the number of photons collected during the qubit reset by optical pumping part of the entanglement generation sequence, averaged over the preceding second, is above a pre-set threshold. If the check does not pass within a matter of seconds, we pause the experiment and scan the laser frequency to find back the qubit reset transition frequency.
    \item Check that the number of photons collected during the spin-photon entanglement part of the entanglement generation sequence, averaged over the preceding second, is above a pre-set threshold. If the check does not pass within a matter of seconds, we pause the experiment and scan the bias voltage of the setup.
\end{itemize}

\begin{figure}
	\centering
	\includegraphics[width=0.5\linewidth]{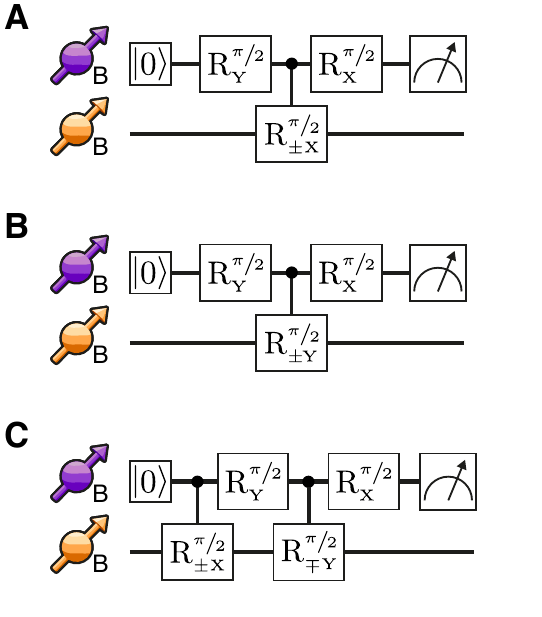}
	\caption{\label{fig:supp_nuclear-spin-control}Memory qubit readout sequences. (A-C) Readout sequences of the nuclear spin memory qubit expectation values for $\hat X, \hat Y, \hat Z$ via the communication qubit. The controlled rotations are to be read as follows: $\text{R}_{\text{\tiny\ensuremath \pm}\text{\tiny X}}^{\sfrac{\pi}{2}}$ is a rotation of the memory qubit around the X axis with an angle of $\pi / 2$ if the communication qubit is in $\ket 0$, and with an angle of $- \pi / 2$ if the communication qubit is in $\ket 1$.}
\end{figure}

\begin{figure}
	\centering
	\includegraphics[width=\linewidth]{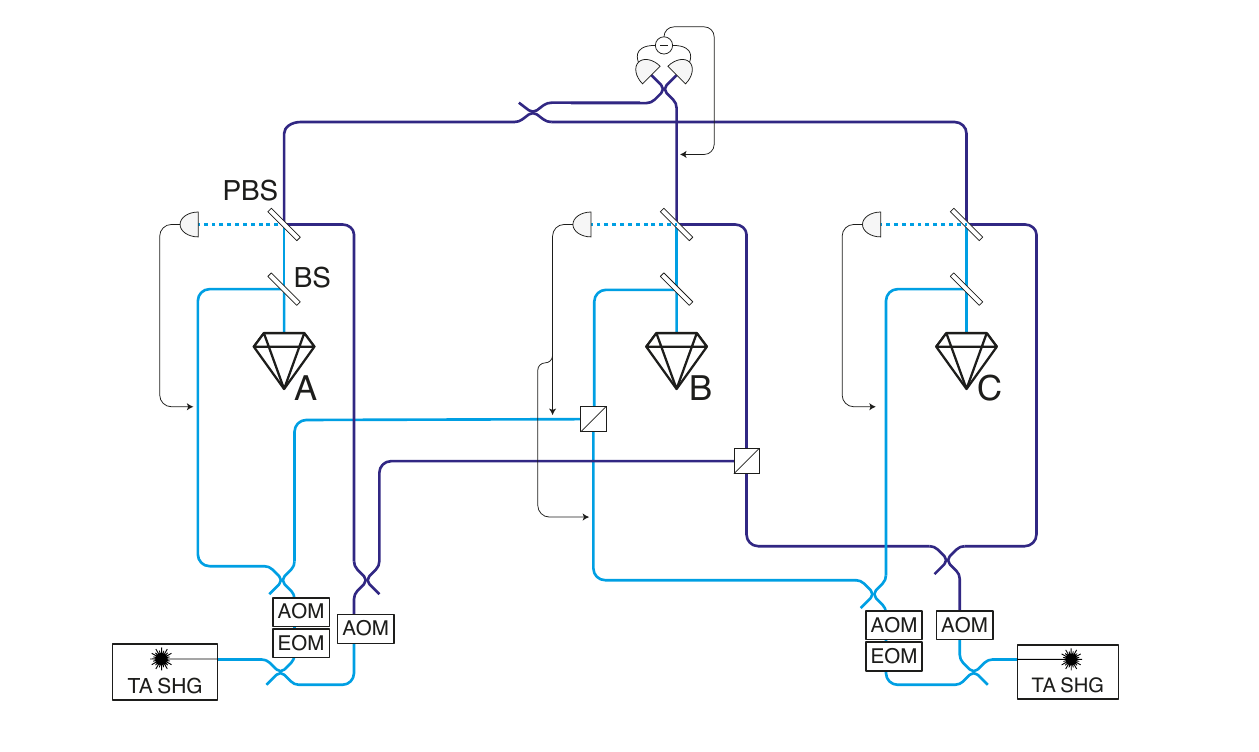}
	\caption{\label{fig:supp_phase_diagram_total}Diagram of the entire layout. Shown are the paths used by the excitation laser (solid light-blue lines) and the phase light (solid dark-blue lines), which has a frequency offset of $\approx \SI{10}{\mega\hertz}$ with respect to the excitation laser. The frequency offset is generated using different frequency modulation set-points for the acousto-optic modulators (AOM) in the excitation and phase path respectively.}
\end{figure}

\begin{figure}
	\centering
	\includegraphics[width=\linewidth]{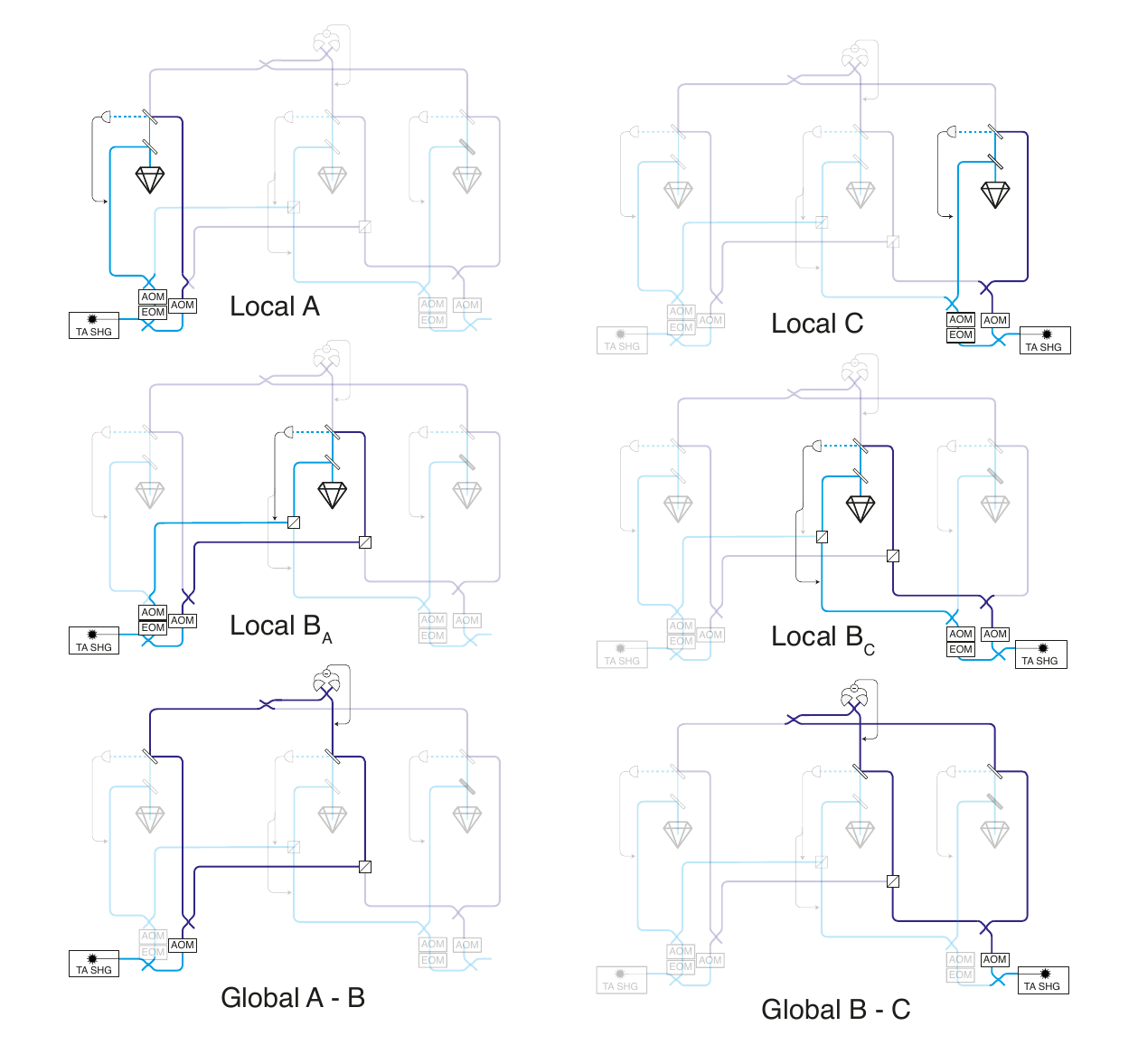}
	\caption{\label{fig:supp_phase_diagram_sub}Diagrams of the six  interferometers in which the optical set-up is divided. For the local interferometers, the heterodyne beat signal (dashed light-blue lines) is measured, compared to an electronic reference signal and feedback is applied to the optical path via piezo-electric mounted mirrors. For the global interferometers, the interference is measured by the single photon detectors. The detectors are shared for the two entanglement links using a beam-splitter that combines the photons from Alice with photons from Charlie. A feedback signal is applied to a fiber stretcher which is also shared by the two global interferometers.}
\end{figure}

\begin{figure}
	\centering
	\includegraphics[width=0.8\linewidth]{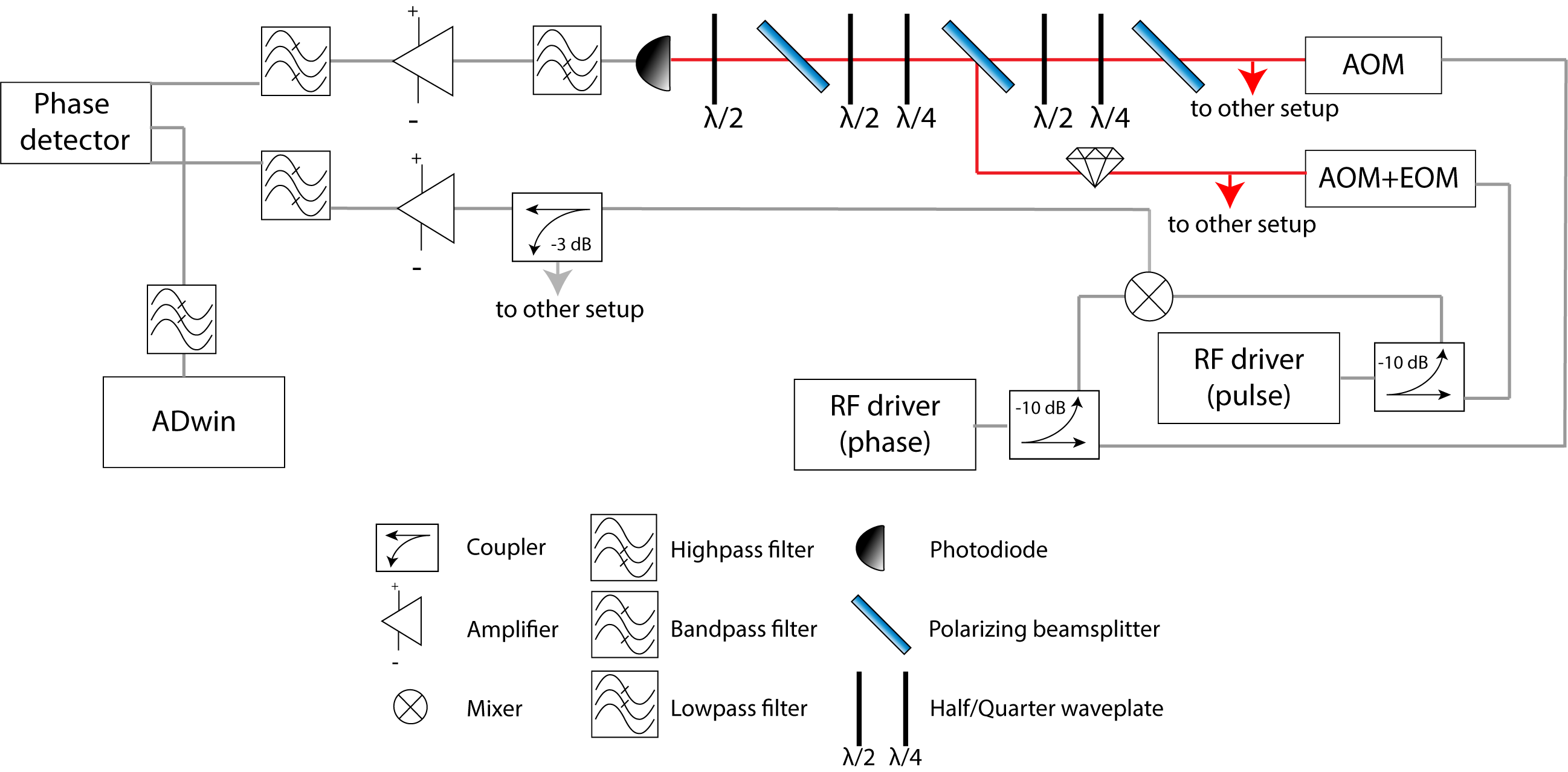}
	\caption{\label{fig:supp_phase_two_node}Diagram of the electronics and optics for a local interferometer using a heterodyne phase detection scheme. Both the electronic reference signal and the excitation light are shared with another setup.}
\end{figure}

 \begin{figure}
 	\centering
 	\includegraphics[width=\linewidth]{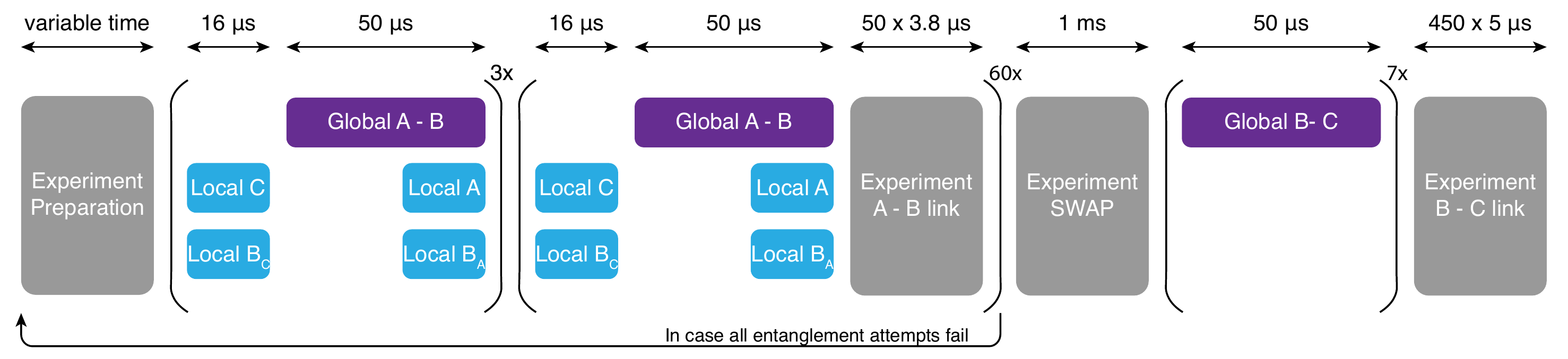}
 	\caption{\label{fig:supp_phase_timings}Overview of the timings related to the phase stabilization. Experimental time (grey blocks) is interleaved with phase stabilization cycles, which include a phase measurement and a feedback. The subscript to B indicates which light is used, either from setup A or C. The local phase stabilization of A and $\mathrm{B_A}$ and the global phase stabilization A-B can be performed at the same time since they use the same light sources.}
 \end{figure}

\begin{figure}
\centering
\begin{minipage}{0.3\linewidth}
	\includegraphics[width=\linewidth]{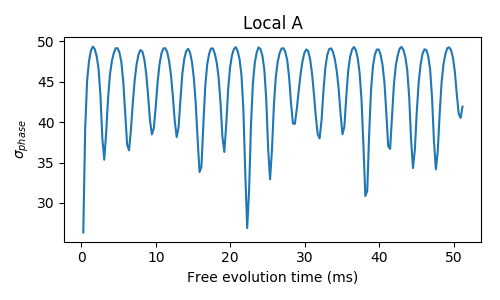}
	\label{fig:free_evolv_lt5}
\end{minipage}
\begin{minipage}{0.3\linewidth}
	\includegraphics[width=\linewidth]{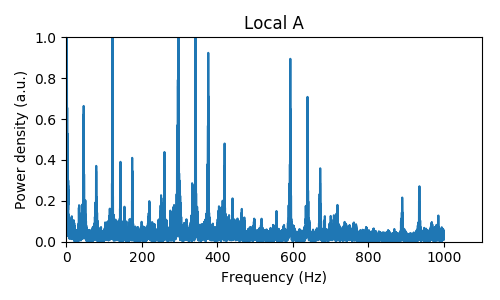}
	\label{fig:spec_lt5}
\end{minipage}
\begin{minipage}{0.3\linewidth}
	\includegraphics[width=\linewidth]{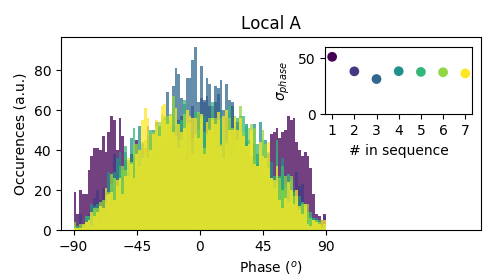}
	\label{fig:sweep_p_lt5}
\end{minipage}
\begin{minipage}{0.3\linewidth}
	\includegraphics[width=\linewidth]{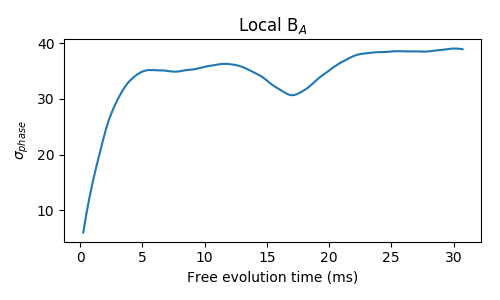}
	\label{fig:free_evolv_lt3a}
\end{minipage}
\begin{minipage}{0.3\linewidth}
	\includegraphics[width=\linewidth]{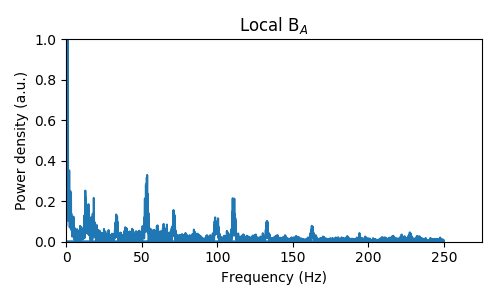}
	\label{fig:spec_lt3a}
\end{minipage}
\begin{minipage}{0.3\linewidth}
	\includegraphics[width=\linewidth]{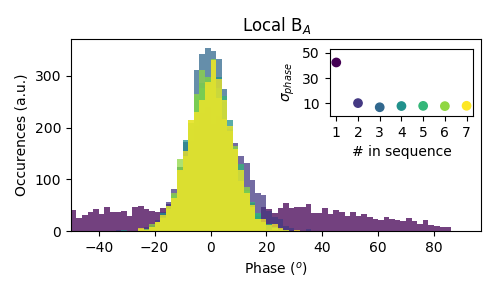}
	\label{fig:sweep_p_lt3a}
\end{minipage}

\begin{minipage}{0.3\linewidth}
	\includegraphics[width=\linewidth]{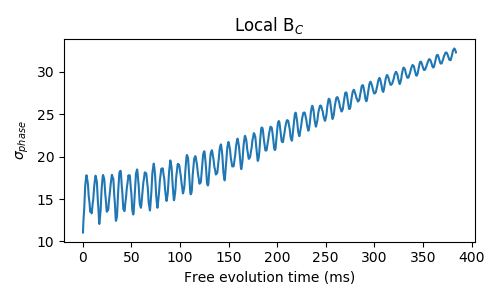}
	\label{fig:free_evolv_lt3b}
\end{minipage}
\begin{minipage}{0.3\linewidth}
	\includegraphics[width=\linewidth]{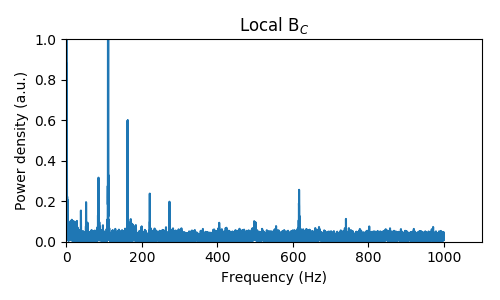}
	\label{fig:spec_lt3b}
\end{minipage}
\begin{minipage}{0.3\linewidth}
	\includegraphics[width=\linewidth]{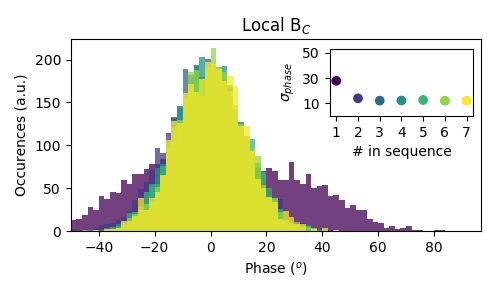}
	\label{fig:sweep_p_lt3b}
\end{minipage}

\begin{minipage}{0.3\linewidth}
	\includegraphics[width=\linewidth]{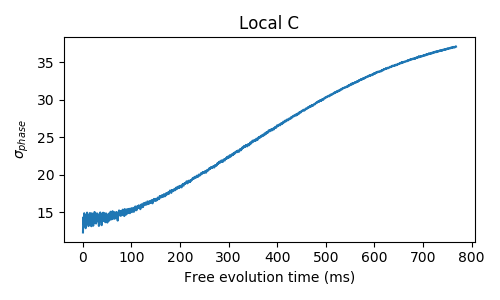}
	\label{fig:free_evolv_lt4}
\end{minipage}
\begin{minipage}{0.3\linewidth}
	\includegraphics[width=\linewidth]{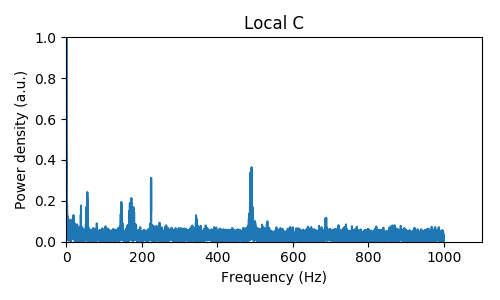}
	\label{fig:spec_lt4}
\end{minipage}
\begin{minipage}{0.3\linewidth}
	\includegraphics[width=\linewidth]{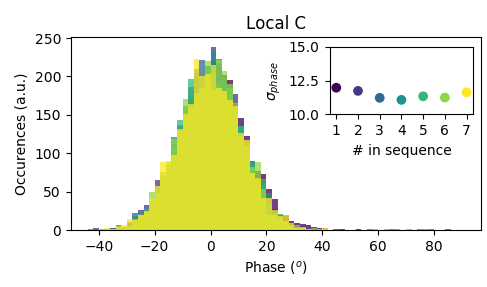}
	\label{fig:sweep_p_lt4}
\end{minipage}

\begin{minipage}{0.3\linewidth}
	\includegraphics[width=\linewidth]{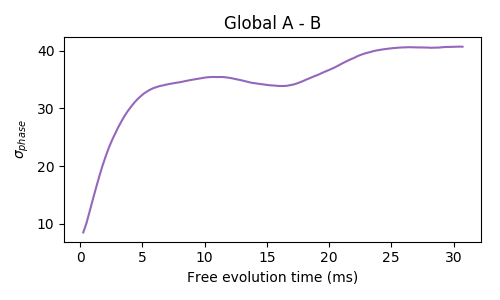}
	\label{fig:free_evolv_gps_ab}
\end{minipage}
\begin{minipage}{0.3\linewidth}
	\includegraphics[width=\linewidth]{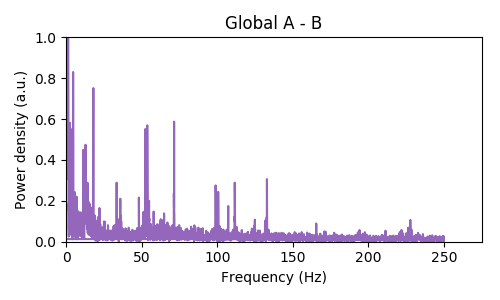}
	\label{fig:spec_gps_ab}
\end{minipage}
\begin{minipage}{0.3\linewidth}
	\includegraphics[width=\linewidth]{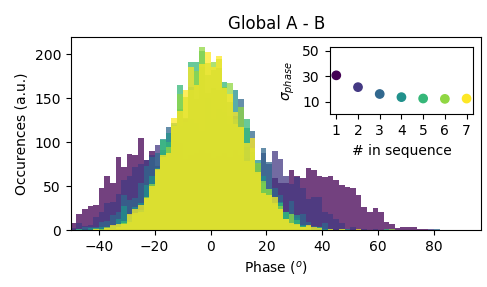}
	\label{fig:sweep_p_gps_ab}
\end{minipage}

\begin{minipage}{0.3\linewidth}
	\includegraphics[width=\linewidth]{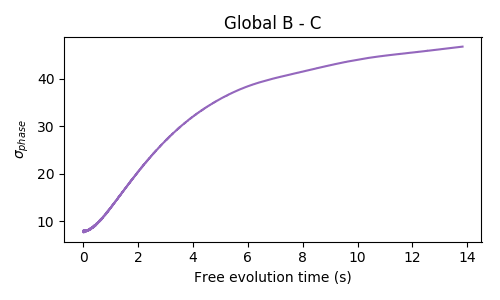}
	\label{fig:free_evolv_gps_bc}
\end{minipage}
\begin{minipage}{0.3\linewidth}
	\includegraphics[width=\linewidth]{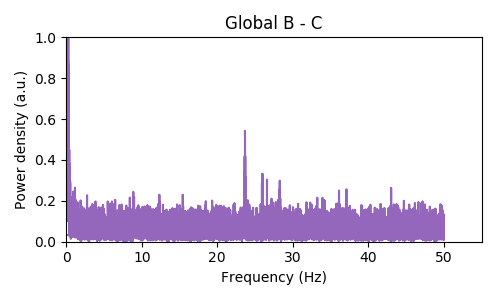}
	\label{fig:spec_gps_bc}
\end{minipage}
\begin{minipage}{0.3\linewidth}
	\includegraphics[width=\linewidth]{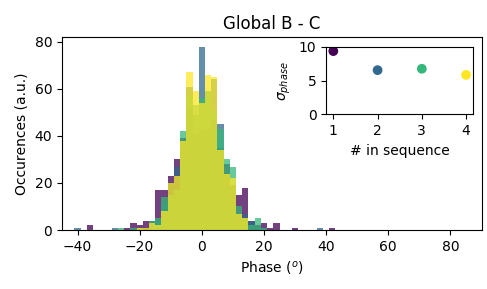}
	\label{fig:sweep_p_gps_bc}
\end{minipage}
\caption{\label{fig:supp_phase_results}Characterization of the phase stabilization of all six interferometers. (Left) Standard deviation of the measured phase while changing the free evolution time. (Center) Frequency spectrum of the measured noise. (Right) Phase distribution for the different rounds of phase stabilization. (Insets) Standard deviation of the phase per stabilization round.}
\end{figure}

\begin{figure}
	\centering
	\includegraphics[width=\textwidth]{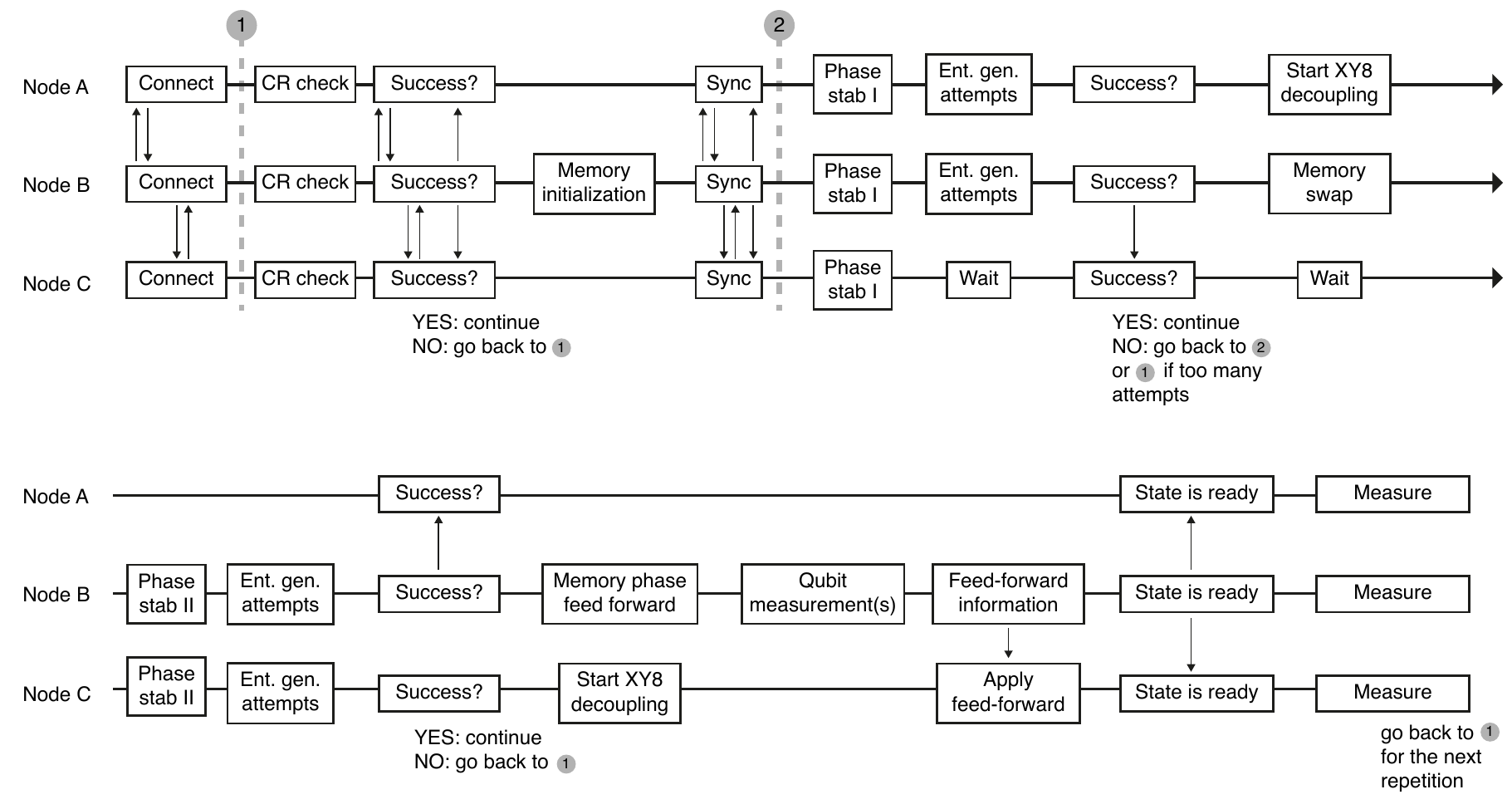}
	\caption{\label{fig:supp_flowchart}Flowchart of the network protocols demonstrations. The micro-controllers of the three setups exchange information to synchronize the experimental sequence and apply feed-forward operations. Vertical lines between the nodes represent communication steps.}
\end{figure}
\begin{sidewaystable}[hb]
\centering
\caption{\label{tab:supp_equipment}Experimental equipment used in the three nodes. Non-listed equipment is unchanged from previous experiments and identical for the three setups.}
\begin{tabular}{|p{3.5cm}|p{6cm}|p{6cm}|p{6cm}|}
\hline
& Alice & Bob & Charlie\\ \hline\hline
Cryostat & attocube attoDRY800 & Montana Instruments Cryostation S50 & Montana Instruments Cryostation S50 \\ \hline
Positioner & Sample on attocube xyz stack & Microscope objective on PI P-615 & Microscope objective on PI P-615 \\ \hline
Micro-controller & J\" ager ADwin-Pro II T12 & J\" ager ADwin-Pro II T12 & J\" ager ADwin-Pro II T12\\ \hline
Arbitrary Waveform Generator (AWG) & Tektronix AWG5014 & Tektronix AWG5014C & Tektronix AWG5014C \\ \hline
MW source & R\&S SGS100A SGMA - up to \SI{6}{\giga\hertz} & R\&S SGS100A SGMA - up to \SI{12.75}{\giga\hertz} & R\&S SGS100A SGMA - up to \SI{6}{\giga\hertz} \\ \hline
MW amplifier & AR 40S1G4 & AR 25S1G4A & AR 40S1G4 \\ \hline
Entangling and qubit readout laser & Toptica TA-SHG pro \SI{637}{nm} & Toptica DL pro \SI{637}{nm} & Toptica TA-SHG pro \SI{637}{nm} \\ \hline
Qubit reset laser & Toptica DL pro \SI{637}{nm} &
\begin{tabular}[c]{@{}c@{}}
    1 - Toptica TA-SHG pro \SI{637}{nm}\\ \\
    2 - New Focus Velocity \SI{637}{nm}
\end{tabular} &
New Focus Velocity \SI{637}{nm} \\ \hline
Charge reset laser & Toptica DL-SHG pro \SI{575}{nm} & Cobolt \SI{515}{nm} & Toptica DL-SHG pro \SI{575}{nm} \\ \hline
EOM & Jenoptik AM635 & & Jenoptik AM635 \\ \hline
Deformable mirror & Boston Micromachines 12x12 & Boston Micromachines 12x12 & Boston Micromachines 12x12 \\ \hline
\end{tabular}
\end{sidewaystable}

\begin{table}[hb]
\centering
\caption{\label{tab:supp_budget_epr}Error budget of the generated Bell states and experimental parameters. The error due to the probability that both nodes emit a photon is related to the values of $\alpha$ (see section \ref{sec:supp_model}) and is therefore intrinsic to the protocol. The infidelity contribution for each of the other errors is estimated as if that error were the only other error present, this way one can easily compare the relative effect of the different infidelity sources. When combined we take into account all the errors at the same time.}
\vspace{1em}
\begin{tabular}{|p{6.5cm}|p{3cm}|p{3cm}|}
\hline
Source of infidelity & \multicolumn{2}{c|}{Expected state infidelity}\\
& Alice - Bob & Bob - Charlie \\ \hline\hline
Probability that both nodes emit a photon & \num{6.1e-2} & \num{8.0e-2} \\ \hline
Phase uncertainty & \num{6.0e-2} & \num{1.5e-2} \\
Double excitation & \num{5.5e-2} & \num{7.0e-2} \\
Photon distinguishability & \num{2.4e-2} & \num{2.3e-2} \\
Non-NV and dark counts & \num{5e-3} & \num{5e-3} \\
\hline 
Combined & \num{0.191} & \num{0.186} \\
\hline \hline
Measured $\Psi^+$ infidelity & \num{0.180(5)} & \num{0.192(5)} \\
Measured $\Psi^-$ infidelity & \num{0.189(5)} & \num{0.189(4)} \\
\hline
\hline
Experimental parameters & Alice - Bob & Bob - Charlie\\ 
\hline
$p_\mathrm{det}$ & \begin{tabular}[c]{@{}c@{}}
    $p^\mathrm{det}_A = \num{3.6e-4}$\\
    $p^\mathrm{det}_B = \num{4.4e-4}$
\end{tabular} &
\begin{tabular}[c]{@{}c@{}}
    $p^\mathrm{det}_B = \num{4.2e-4}$\\
    $p^\mathrm{det}_C = \num{3.0e-4}$
\end{tabular} \\
\hline
$\alpha$ & \begin{tabular}[c]{@{}c@{}}
    $\alpha_A = \num{0.07}$\\
    $\alpha_B = \num{0.05}$
\end{tabular} &
\begin{tabular}[c]{@{}c@{}}
    $\alpha_B = \num{0.05}$\\
    $\alpha_C = \num{0.10}$
\end{tabular} \\
\hline
$p_\mathrm{dc}$ & \num{1.5e-7} & \num{1.5e-7}\\
\hline
Visibility $V$ & \num{0.90} & \num{0.90}\\
\hline
Phase uncertainty & \ang{30} & \ang{15}\\
\hline
Entanglement attempt duration & \SI{3.8}{\micro\second} & \SI{5.0}{\micro\second}\\
\hline
Probability of double excitation & \num{0.06} & \num{0.08}\\
\hline
\end{tabular}
\end{table}

\begin{table}[hb]
\centering
\caption{\label{tab:supp_fit_results}Fit results for the curves displayed in Figure 3 of the main text with and without entanglement generation (Ent. Gen.). See section~\ref{sec:supp_memory_lifetime} for details on the fitting function.}
\vspace{1em}
\begin{tabular}{|p{1cm}|p{3.5cm}|p{3.5cm}|}
\hline
& With Ent. Gen. & Without Ent. Gen.\\ \hline\hline
$N_{1/e}$ & \num{1843(32)} & \num{2042(36)} \\
$n$ & \num{1.37(5)} & \num{1.61(6)} \\
$A$ & \num{0.895(6)} & \num{0.885(6)} \\
\hline
\end{tabular}
\end{table}

\begin{table}[hb]
\centering
\caption{\label{tab:supp_budget_ghz}Error budget of the generated GHZ states. Initially each error is estimated as if it were the only source of error. When combined we take into account all the errors at the same time.}
\vspace{1em}
\begin{tabular}{|p{5.5cm}|p{5cm}|}
\hline
Source of infidelity & Expected GHZ state infidelity\\ \hline\hline
$\Psi_{AB}$ state infidelity & \num{0.191}\\
$\Psi_{BC}$ state infidelity & \num{0.186}\\
Nuclear-spin dephasing noise & \num{2.8e-2}\\
Nuclear-spin depolarising noise & \num{8.3e-2}\\
Feed-forward errors & \num{6e-3}\\
\hline 
$\Psi_{AB}$ and $\Psi_{BC}$ combined &  \num{0.337}\\
Combined &  \num{0.406}\\
\hline \hline
Measured GHZ infidelity & \num{0.462(18)} \\
\hline
\end{tabular}
\end{table}

\begin{table}[hb]
\centering
\caption{\label{tab:supp_budget_swapping}Error budget of the generated Alice-Charlie states. Initially each error is estimated as if it were the only source of error. When combined we take into account all the errors at the same time. Errors reported for different Bell state measurement (BSM) results.}
\vspace{1em}
\begin{tabular}{|p{7.5cm}|p{5cm}|}
\hline
Source of infidelity & Expected $\Phi_{AC}$ state infidelity\\ \hline\hline
$\Psi_{AB}$ state infidelity & \num{0.191}\\
$\Psi_{BC}$ state infidelity & \num{0.186}\\
Nuclear-spin dephasing noise & \num{2.8e-2}\\
Nuclear-spin depolarising noise & \num{8.2e-2}\\
Feed-forward errors (00 BSM result) & \num{1.3e-2}\\
Feed-forward errors (any BSM result) & \num{7.5e-2}\\
\hline 
Combined (00 BSM result) & \num{0.398}\\
Combined (any BSM result) & \num{0.428}\\
\hline \hline
Measured $\Phi_{AC}$ infidelity (00 BSM result) & \num{0.413(28)} \\
Measured $\Phi_{AC}$ infidelity (any BSM result) & \num{0.449(13)} \\
\hline
\end{tabular}
\end{table}

\end{document}